\documentclass[12pt]{article}
\pdfoutput=1
\usepackage{putex}

\usepackage{amsmath, mathtools, amssymb}
\usepackage{graphicx,hyperref}
\usepackage{caption}
\usepackage{subcaption}
\usepackage{epstopdf}
\usepackage{enumerate}
\usepackage{cite}
\usepackage{tensor}
\usepackage{slashed}
\usepackage[aligntableaux=center]{ytableau}

\numberwithin{equation}{section}

\newcommand {\be} {\begin {equation}}
\newcommand {\ee} {\end {equation}}

\newcommand {\bes} {\begin {equation*}}
\newcommand {\ees} {\end {equation*}}

\newcommand{\CP}{\mathbb{CP}}
\newcommand{\Z}{\mathbb{Z}}

\newcommand{\R}{\mathbb{R}}
\newcommand{\C}{\mathbb{C}}

\newcommand{\beq}{\begin{equation}}
\newcommand{\eeq}{\end{equation}}

\def\<{\langle}
\def\>{\rangle}

\newcommand{\cH}{\ensuremath{\mathcal{H}}}

\newcommand{\cO}{\ensuremath{\mathcal{O}}}

\newcommand{\dd}{\mathrm{d}}

\newcommand{\fs}{\mathbb{FS}}

\newcommand\vertarrowbox[3][6ex]{%
	\begin{array}[t]{@{}c@{}} #2 \\
		\left\downarrow\vcenter{\hrule height #1}\right.\kern-\nulldelimiterspace\\
		\makebox[0pt]{#3}
	\end{array}%
}

\begin{document}

\institution{SCGP}{Simons Center for Geometry and Physics,\cr Stony Brook University, Stony Brook, NY 11794-3636, USA}

\title{Ising BCFT from Fuzzy Hemisphere}
\authors{Mykola Dedushenko\worksat{\SCGP}}

\abstract{We extend the recently introduced fuzzy sphere technique for the 3d Ising CFT to the case of boundary CFT (BCFT) using the fuzzy hemisphere. This allows to study conformal boundary conditions, and we investigate the three boundary universality classes of the 3d Ising CFT. These include the ordinary, extraordinary, and special boundary conditions. We identify the qualitative signatures of boundary conformal invariance, and in the cases of ordinary and extraordinary boundary conditions, obtain numerical estimates for the spectrum of low-lying boundary primaries. For the special boundary conditions, we are only able to see their qualitative features, while reliable estimates of the scaling dimensions of boundary primaries are left for the future work.}

\date{}

\maketitle

\tableofcontents

\section{Introduction}\label{sec:intro}
Conformal field theories (CFT) play fundamental role in physics, with applications ranging from quantum and classical statistical mechanics, to string theory, to formal QFT. Of special relevance are 2d CFTs, due to their role in string theory and 2d statistical systems, and 3d CFTs, as they describe thermal phase transitions in 3d materials (magnets, water, etc). A lot is known analytically in two dimensions (Euclidean 2d or Minkowskian $(1+1)$d) thanks to the infinite-dimensional Virasoro symmetry and its extensions \cite{Belavin:1984vu}. Similar analytical tools are not yet available for higher-dimensional CFTs. They are studied either perturbatively, or using numerical techniques, which, until recently, included the conformal bootstrap \cite{Poland:2018epd} (capable of providing rigorous numerical bounds), and techniques to study CFT by probing critical points of known many-body systems (via Monte Carlo simulations, exact diagonalization (ED), density matrix renormalization group (DMRG)), see for instance \cite{Deng:2005dh,Cosme:2015cxa,ferrenberg2018pushing,Meneses:2018xpu,Lao:2023zis}.

The last two years have seen an emergence of a new numerical technique in 3d CFT -- the fuzzy sphere method \cite{Zhu:2022gjc,Hu:2023xak,Han:2023yyb}. To be more precise, this method gives an approach to constructing a many-body system, whose critical point (quantum critical point, in fact,) realizes the $(2+1)$d CFT in question. Once such a system is given, it can be solved using the standard tools, such as ED and DMRG in \cite{Zhu:2022gjc} or even Monte Carlo \cite{Hofmann:2023llr}, providing approximate CFT data once tuned to the quantum critical point.

In 3d CFT, local operators on $\R^3$ are identified, via the state-operator correspondence (see, e.g., \cite{Rychkov:2016iqz},) with the states on $S^2\times \R$. This implies that there exists a unique way to put a 3d conformal theory on $S^2\times \R$. A 3d CFT on $S^2\times \R$ may be thought of as the quantum critical point of some underlying (usually finite-dimensional) quantum-mechanical system on $S^2\times \R$. Here, however, one may encounter difficulties: Even if we know the necessary system in flat space, on $S^2\times \R$ one may find ambiguities. The main example to keep in mind is of course the Ising model, which is a system of variables $\sigma_{\vec x}=\pm1$ with the nearest neighbor interaction $H=-J\sum_{\langle \vec{x}\vec{y} \rangle} \sigma_{\vec x}\sigma_{\vec y}$, usually defined on the square lattice. Placing it on $S^2$ requires modifying the lattice, as the square lattice would not work for topological reasons. Furthermore, states in a general 3d CFT on $S^2$ come in the multiplets of the rotational group $SO(3)$ (or, more generally, ${\rm Spin}(3) = SU(2)$ in a fermionic theory). A lattice on $S^2$ would break this rotational group to a finite subgroup, which makes it harder to identify the $SU(2)$ multiplets at the CFT point. (Using lattices, however, has been successfully attempted before \cite{MWeigel_2000,Deng:2002ea,Lao:2023zis,Brower:2024otr}.)

In the fuzzy sphere method, instead of replacing $S^2$ by a lattice, one uses the non-commutative or fuzzy approximation to $S^2$ known as the fuzzy sphere \cite{Madore:1991bw}, which will be denoted as $\fs^2_N$. Here $N$ is the ``UV cutoff'' parameter, such that functions on the fuzzy sphere are represented by the $N\times N$ matrices, and $\frac1{N}$ is the non-commutativity parameter. In general, $N$ is any positive integer, but in this paper it is assumed to be even. Just like the lattice, fuzziness achieves the main goal of any UV regularization, -- replaces the infinite-dimensional algebra of functions on $S^2$ by the finite-dimensional algebra (which is the algebra of $N\times N$ matrices in this case). Unlike the lattice, it has a significant advantage -- it preserves the full rotational $SU(2)$ symmetry. Extra symmetry means more ``protection'' from the unwanted ``finite-size'' effects (plus, of course, the identification of $SU(2)$ multiplets becomes manifest now.)  In the past, there have been attempts to use the non-commutativity as a UV regulator in QFT \cite{Nekrasov:1998ss,Gubser:2000cd,Douglas:2001ba,Szabo:2001kg,Seiberg:1999vs,Minwalla:1999px,Matusis:2000jf,Chu:2001xi,Vaidya:2003ew}, which, however, encountered various unusual phenomena, such as the UV/IR mixing and the ``noncommutative anomaly''.

The intuitive idea behind the novel approach of \cite{Zhu:2022gjc} is to use the auxiliary degrees of freedom, --- ``electrons'' -- that live on the fuzzy $\fs^2_N$. These non-commutative degrees of freedom are used as a ``substrate'', a sort of replacement of the lattice. Then the actual commutative quantum system is built on top of this substrate. For example, in \cite{Zhu:2022gjc}, one endows electrons with the Ising isospin $\pm\frac12$, and turns on the appropriate interactions consistent with the Ising universality class. By tuning them to the quantum critical point and diagonalizing the Hamiltonian, the authors of \cite{Zhu:2022gjc} were able to recover the conformal data of the 3d Ising CFT. In \cite{Zhou:2023qfi}, electrons were given other internal, or ``flavor'', degrees of freedom, and interactions were tuned to the quantum critical point describing the $SO(5)$ deconfined phase transition.

The precise identification of the quantum-mechanical system on $\fs^2_N$, however, is not very systematic yet, and remains somewhat of an art form at the moment. The situation is not too different from the typical state of affairs in QFT, when we try to study the fixed points of the renormalization group (RG) flow. One fixes symmetries, a certain set of fields, and then considers all possible interactions consistent with the said symmetries. Tuning relevant interactions generically selects a CFT at the RG fixed point in the IR, while irrelevant interactions die off in the strict IR limit. In practice, we perform computations (in our case, numerically) at finite energies, not in the strict IR limit. Thus, the irrelevant interactions contribute and remain important. One ends up with the trade-off problem: On the one hand, including and carefully tuning more and more irrelevant interactions is going to improve the numerical precision of the final answers. On the other hand, adding more free parameters makes the model less predictive, and fine tuning too many parameters can be just too hard, especially if diagonalizing Hamiltonian at each set of parameters is computationally expensive.

For example in \cite{Zhu:2022gjc}, the authors tune two parameters: The spin-spin interaction strength $V_0$ and the transverse magnetic field $h$. The space of such parameters is divided into two regions, the ordered and the disordered phase, with the CFT living at the boundary. Thus only one combination of these parameters, roughly, measuring distance to the phase boundary in the $(V_0, h)$ space,  corresponds to the relevant operator, which is of course the mass operator $\sigma^2$ in the Landau-Ginzburg description of this phase transition. The other combination of parameters is some irrelevant coupling, telling us precisely at what point of the phase boundary in the $(V_0, h)$ space we sit. In \cite{Zhu:2022gjc}, this irrelevant coupling was tuned to the value that minimizes the ``finite-size'' (or $1/\sqrt{N}$) effects. In practice this means that the spectrum of energies looks ``more conformal'', and one can easily identify the conformal multiplets by observing that the energy gaps between states in the multiplet are integers to high enough precision. In theory, it should also be possible to map out the precise correspondence between the isospin couplings and various CFT operators (like, e.g., in \cite{Zou:2019dnc}).

Again, such tuning of the relevant couplings, and of a small number of irrelevant ones, is typical in effective field theory (EFT). Then what is so special about the fuzzy sphere approach? In a sense, only that it manifestly preserves the spatial $SU(2)$ rotations. The $SU(2)$ symmetry, on the one hand, restricts possible interactions that can appear in the effective action; on the other hand, it protects the spectrum: Operators $\cO_{-j}, \cO_{-j+1}, \dots, \cO_j$ in the same $SU(2)$ multiplet are all forced to have the same scaling dimension, whereas without the $SU(2)$ symmetry, each of them could receive independent finite-size corrections, possibly of different signs.

The method has been successfully applied to study the bulk CFT \cite{Zhu:2022gjc,Hu:2023xak,Han:2023yyb}, its line defects \cite{Hu:2023ghk,Zhou:2023fqu,Cuomo:2024psk}, and the F-function \cite{Hu:2024pen}, all in the case of 3d Ising model, with the exception of \cite{Zhou:2023qfi,Chen:2024jxe}. One currently missing application is to surfaces and boundaries, i.e., to the setting of boundary CFT (BCFT). When a CFT is defined on the half-space $\R^2\times \R_>$, a Weyl transformation allows to put it on $HS^2\times \R$, where now $HS^2$ is a hemisphere, -- this is the radial quantization in BCFT. Thus all we need to do is replace the fuzzy $\fs^2_N$, by the fuzzy hemisphere. Of course on the hemisphere, the $SU(2)$ is broken to its maximal torus $U(1)$ of the hemisphere rotations. Thus, one of the superpowers of $\fs^2_N$ is lost, and larger finite-size effects are expected. Nevertheless, the surviving $U(1)$ is still better than the finite group that would be preserved by the lattice. Therefore, we expect that on the hemisphere, we should still be able to get some mileage out of the fuzzy techniques.

But what is the fuzzy hemisphere, anyways? The fuzzy sphere \cite{Madore:1991bw}, as we will review later, is the $N$-dimensional representation of $SU(2)$, or spin-$s$ irrep, where $s=\frac{N-1}{2}$. In the large-$N$ limit, the sphere of area $\sim N$ emerges as a subvariety in the representation space $\C^N$ \cite{Ded:fuzzy}. It consists of $N$ evenly spaced orbitals, or parallels (the eigenstates for $U(1)\subset SU(2)$), each having thickness of order $1/\sqrt{N}$. Since in the large-$N$ limit such orbitals are localized, one possible way to define the fuzzy hemisphere is by only keeping the orbitals in one hemisphere, the associated weight subspace being $\C^{N/2}$ (which is why $N$ is even). In this paper, we choose to keep orbitals in the Southern hemisphere (as in Figure \ref{fig:true-hemi}), and use that as a definition of the fuzzy hemisphere. Another approach to our problem is to keep the full fuzzy sphere $\fs^2_N$, however, only make degrees of freedom (``electrons'' and their isospins) dynamical in the Southern hemisphere, while those in the Northern hemisphere are frozen to the state of our choice. We will alternate between these two approaches in what follows.

We will study the 3d Ising BCFT, the first example of applying the fuzzy sphere techniques to BCFT. It is known \cite{Cardy:1996xt,Diehl:1986,Diehl:1998mh,Deng:2005dh,Liendo:2012hy,Gliozzi:2015qsa} that the 3d Ising model has three types of boundary conditions, see also \cite{Metlitski:2020cqy,Padayasi:2021sik,Toldin:2021kun,Trepanier:2023tvb}. They are determined by whether the boundary goes through the order/disorder phase transition at the higher or lower temperature than the bulk. If the boundary remains disordered while the bulk goes through such a transition (e.g., because the boundary spins are very weakly coupled), the corresponding conformal boundary condition is, somewhat unfortunately, called ``ordinary''. If the boundary is ordered and spontaneously breaks the Ising $\Z_2$ symmetry, the boundary condition is called extraordinary. In this case, the theory decomposes into two sectors related by the broken $\Z_2$, both known as the ``normal'' boundary classes (thus ``extraordinary'' and ``normal'' boundary conditions are equivalent at the level of local operators, but might have different extended operators). Finally, if the bulk and boundary go through the order/disorder transition simultaneously, the corresponding conformal boundary condition is known as ``special''. Both ``ordinary'' and ``special'' boundary conditions are $\Z_2$-preserving, and in the Landau-Ginzburg description, they are realized by the Dirichlet and Neumann boundary conditions on the order parameter $\sigma$, respectively \cite{Liendo:2012hy}. The special boundary supports one $\Z_2$-odd boundary primary that we denote $\hat\sigma$, and a $\Z_2$-even one that we suggestively denote $:\hat\sigma^2:$. The ordinary boundary only has one $\Z_2$-odd boundary primary, and the extraordinary case supports no relevant operators.

In the current paper, we study these three universality classes of boundary conditions. In Section \ref{sec:fuzzy-review}, we start by reviewing some basics of the fuzzy sphere, including the approach of \cite{Zhu:2022gjc} to 3d Ising CFT. In Section \ref{sec:fuzzy-BCFT}, we discuss the 3d Ising BCFT, and extend the fuzzy techniques to this case. In Section \ref{sec:num} we describe our numerics. We start with some general comments in Section \ref{sec:sorcery} on how we minimize the finite-size effects. In Sections \ref{sec:z2-break} and \ref{sec:z2-preserve}, we discuss the extraordinary and ordinary boundary classes, where the numerical results are quite conclusive. Then in Section \ref{sec:special}, we give partial results on the special boundary class: While we are able to see some expected qualitative features of such a boundary, we are not able to give reliable estimate of the scaling dimension of the second boundary primary scalar. We note that the current literature still appears to lack such an estimate. Then we conclude with possible future directions. We note that we do not attempt to estimate errors in our analysis. However, in the case of ordinary and extraordinary boundary classes, the numerical answers match the expected values with a fairly good precision.

\section{Landau quantization and CFT}\label{sec:fuzzy-review}
Let us start by reviewing the fuzzy sphere regularization in 3d CFT introduced in \cite{Zhu:2022gjc}. In radial quantization, we put a 3d CFT on the product manifold $S^2\times\R$, and identify the Hilbert space $\cH[S^2]$ with the space of local operators. In particular, the energy spectrum of the Hamiltonian is
\begin{equation}
E_n = E_0 + \alpha \Delta_n,
\end{equation}
where $E_0$ is known as the Casimir energy, $\Delta_n$ are scaling dimensions, and $\alpha = \frac{c}{R}$ is a proportionality constant. The numbers $\Delta_n$ are of primary interest, though $E_0$ is also an interesting quantity \cite{Assel:2015nca,Luo:2022tqy}.

Now suppose that our CFT occurs at the quantum critical point of some many-body system \cite{Polyakov:1970xd,Cardy:1996xt}. One can imagine, say, having a lattice system on $S^2$ subject to quantum evolution, which is described by a $(2+1)$d CFT at large distances, once the parameters are tuned to a quantum critical point. An obvious difficulty here is that a lattice, normally defined in flat space, requires modifications to be placed on $S^2$, which introduces lattice defects. Also, the lattice breaks the $SO(3)$ rotations down to some finite subgroup. The authors of \cite{Zhu:2022gjc} came up with a new way to circumvent these issues by using the fuzzy sphere regularization instead of the lattice one.

Let us recall (see the review in \cite{Ded:fuzzy}) that the fuzzy sphere consists of a vector space identified with the global holomorphic sections of line bundle $\cO(n)$ over $\CP^1$:
\begin{equation}
V_n = \C^{n+1} \cong H^0(\CP^1, \cO(n)),
\end{equation}
with the functions on $\CP^1$ approximated by the $(n+1)\times (n+1)$ matrices. Such an approximation is what we called $\fs^2_N$ in the introduction, with
\begin{equation}
N=n+1,
\end{equation}
More concretely, we think of vectors (states) $\Psi\in V_n$ as degree $n$ polynomials in the projective coordinates on $\CP^1$:
\begin{equation}
\Psi = \sum_{k=0}^n c_k z_0^k z_1^{n-k}.
\end{equation}
Then functions on $\fs^1_N$ are represented by the degree-preserving differential operators acting on $V_n$:
\begin{equation}
\label{functions-fuzzy}
F = \sum_{\substack{d_1+d_2=n\\d_3+d_4=n}} c_{d_1,d_2,d_3,d_4} z_0^{d_1} z_1^{d_2} \frac{\partial^{d_3}}{\partial z_0^{d_3}} \frac{\partial^{d_4}}{\partial z_1^{d_4}}.
\end{equation}
These in the $n\to\infty$ limit become analytic function on $\CP^1$ (represented as $U(1)$-invariant functions in $z_0, \bar{z}_0, z_1, \bar{z}_1$, since the functions on $\CP^1$ can be lifted to $U(1)$-invariant functions on $S^3$ via the Hopf fibration, and the $S^3$ is embedded as $|z_0|^2 + |z_1|^2=1$ inside $\C^2$). One of the most appealing features of the fuzzy sphere is that the differential operators \eqref{functions-fuzzy} form a closed algebra of $N\times N$ matrices (which approximates the algebra of analytic function on $\CP^1$ in the $N\to \infty$ limit).

Assuming $|z_0|^2 + |z_1|^2=1$ and quotienting out the $U(1)$, we can write $z_0 = \cos\frac{\theta}{2} e^{i\varphi/2}$ and $z_1 = \sin\frac{\theta}{2} e^{-i\varphi/2}$. This makes it apparent that the basis of holomorphic sections (or fuzzy orbitals) $z_0^k z_1^{n-k}$ is the same thing as the monopole sphere harmonics \cite{Wu:1976ge,Zhu:2022gjc}:
\begin{align}
\text{Basis of } H^0(\CP^1, \cO(n)) &\leftrightarrow \text{fuzzy orbitals } z_0^k z_1^{n-k} = e^{-in \varphi/2 +ik \varphi} \cos^k \frac{\theta}{2} \sin^{n-k}\frac{\theta}{2}\cr
&\leftrightarrow \text{monopole harmonics } \Phi_m = N_m e^{im \varphi} \cos^{\frac{n}{2}+m} \frac{\theta}{2} \sin^{\frac{n}{2}-m}\frac{\theta}{2},
\end{align}
where $m=-\frac{n}{2}, -\frac{n}{2}+1, \dots, \frac{n}{2}$, and the normalization is $N_m=\sqrt{\frac{(2s+1)!}{4\pi (s+m)!(s-m)!}}$. We see that, up to an overall normalization, the monopole harmonics are the same thing as holomorphic sections of $\cO(n)$, i.e., they are not functions on $\CP^1$. On the other hand, Hermitian inner products of such sections, e.g. $\bar{z}_0^m \bar{z}_1^{n-m} z_0^k z_1^{n-k}$, are functions on $\CP^1$, and in the $n\to\infty$ limit they become the usual spherical harmonics forming the complete basis in the space of functions on $\CP^1$.

One well-known way to ``physically realize'' the fuzzy sphere is by the lowest Landau level (LLL) projection, which sometimes goes under the name of Berezin-Toeplitz quantization in the mathematical literature \cite{MaMarinescu-book,MaMarinescu+2012+1+56,schlichenmaier2010berezin}. Namely, consider an electron moving on $S^2$ with $n$ units of uniform magnetic flux. The energy levels of electron then arrange into the Landau bands \cite{greiter2011landau}. Tuning the mass of electron to zero, we effectively project to the LLL, which happens to be the same vector space $V_n$ that is part of the fuzzy sphere construction. Curiously, one can also project to the LLL by tuning the magnetic flux $n\to\infty$. In this limit, however, the noncommutativity (or fuzziness) goes away, and one regains the commutative $S^2$, albeit as a \emph{phase space} of the electron (before the LLL projection, $S^2$ is the \emph{configuration space}).

The approach of \cite{Zhu:2022gjc}, following \cite{Ippoliti:2018ojo}, is to use the electrons on the fuzzy sphere as auxiliary degrees of freedom serving as a substrate, on which one can realize a critical point described by a given CFT. In the Landau-Ginzburg description of the phase transition, the order parameter is going to be a function on $S^2$, not a section, hence one expects it to be bilinear in the electrons. One starts by giving the electrons some internal degrees of freedom and partially filling the LLL (half-filling, in fact, to ensure that the particle-hole conjugation is a symmetry). At this point, the electrons are free. Then the interaction between them is turned on and tuned to a critical point. At the critical point, the CFT data is studied using the numerical diagonalization of the Hamiltonian. The exact CFT is supposed to emerge in the commutative $N\to\infty$ limit. In this procedure, our main enemy are the ``finite-size'' effects due to the finite $N$, which plays the role of area of the fuzzy $S^2$ (and $N^{-1}$ or $n^{-1}=(N-1)^{-1}$ is the ``Planck constant'' controlling the noncommutativity).

Let us focus on the case of 3d Ising CFT, following \cite{Zhu:2022gjc}. Each electron is given the flavor space $\C^2$, i.e., the ``isospin'' $\frac12$ degree of freedom (unlike the actual spin, it does not transform under the spatial rotations and does not couple to the auxiliary magnetic field creating the noncommutativity).\footnote{This isospin is also, coincidentally, the Ising spin in this model.} Thus every fuzzy orbital $z_0^k z_1^{n-k}$ can support precisely two electrons: one isospin up and one isospin down, due to their odd statistics. That is, there are $2N$ total slots for the electrons, where $N=n+1$. We fill precisely one half of them \cite{Ippoliti:2018ojo} by adding $N$ such electrons on $S^2$. The dimension of the corresponding multi-electron Hilbert space is
\begin{equation}
{2N \choose N}.
\end{equation}
The electrons are arranged into the fermionic fields:
\begin{equation}
\psi_{\uparrow\downarrow} = \frac1{\sqrt N}\sum_{m} \Phi^*_m c_{m,\uparrow\downarrow}\, ,\quad \psi^\dagger_{\uparrow\downarrow} = \frac1{\sqrt N} \sum_{m} \Phi_m, c^\dagger_{m,\uparrow\downarrow}\, ,
\end{equation}
where $c_{m,\uparrow}$ annihilates and $c^\dagger_{m,\uparrow}$ creates electrons of spin up at the m'th fuzzy orbital, while $c_{m,\downarrow}$, $c^\dagger_{m,\downarrow}$ does the same to the electrons of spin down. The density of spin up or down is:
\begin{equation}
n_\uparrow = \psi^\dagger_\uparrow \psi_\uparrow,\quad n_\downarrow = \psi^\dagger_\downarrow \psi_\downarrow.
\end{equation}
The authors of \cite{Zhu:2022gjc} then consider the following Hamiltonian on $S^2$:
\begin{equation}
\label{Ham}
H = 4N^2 \int \dd\Omega_a \dd\Omega_b\, U(\Omega_{ab}) n_\uparrow(\Omega_a) n_\downarrow(\Omega_b) - h N \int \dd\Omega\,  n^x(\Omega),
\end{equation}
where $n^x = (\psi^\dagger_\uparrow, \psi^\dagger_\downarrow)\sigma^x \left(\substack{\psi_\uparrow \\ \psi_\downarrow}\right)$ couples to the transverse magnetic field $h$. The density-density interaction is taken to be ultra-local, $U(\Omega_{ab}) = \frac{g_0}{R^2}\delta(\Omega_{ab}) + \frac{g_1}{R^4} \nabla^2 \delta(\Omega_{ab})$. In terms of the creation-annihilation operators, the Hamiltonian reads:
\begin{align}
\label{HamExpl}
H = \sum_{m_1, m_2, m_3, m_4} (V_{m_1,m_2,m_3,m_4} + V_{m_2,m_1,m_4,m_3}) (c^\dagger_{m_1,\uparrow} c_{m_4,\uparrow})(c^\dagger_{m_2,\downarrow}c_{m_3,\downarrow}) \delta_{m_1+m_2, m_3+m_4}\cr
- h \sum_m (c^\dagger_{m,\uparrow} c_{m,\downarrow} + c^\dagger_{m,\downarrow}c_{m,\uparrow}),
\end{align}
where the interaction coefficients are given in terms of Wigner 3j-symbols:
\begin{equation}
\label{V-coeffs}
V_{m_1, m_2, m_3, m_4} = \sum_{\ell=0,1} V_{\ell} (2n-2\ell+1) {{\frac{n}{2}\qquad \frac{n}{2}\qquad\quad n-\ell}\choose{m_1\quad  m_2 \quad  -m_1-m_2}}{{\frac{n}{2}\qquad \frac{n}{2}\qquad\quad n-\ell}\choose{m_4\quad  m_3 \quad  -m_3-m_4}}.
\end{equation}
Here $V_\ell$ are the Haldane pseudopotentials \cite{Haldane:1983xm}, which in the ultra-local case only involve $V_0$ and $V_1$. In \cite{Zhu:2022gjc}, the energy units are chosen so that $V_1=1$, and the model is studied as a function of the spin-spin interaction strength $V_0$ and the transverse magnetic field $h$.

The model has a lot of symmetry:
\begin{enumerate}
	\item $SU(2)$ rotations of $\CP^1$ act on each collection $(c_{-s,\uparrow}, c_{-s+1, \uparrow, \dots, c_{s, \uparrow}})$ or $(c_{-s,\downarrow}, c_{-s+1, \downarrow, \dots, c_{s, \downarrow}})$ in the spin-$s$ representation. Again, the ``isospin'' degree of freedom $\uparrow$, $\downarrow$ is not acted on by the rotations. We could alternatively call it a qubit.
	\item The usual Ising $\Z_2$ symmetry that flips such qubits or isospins.
	\item The particle-hole symmetry, which is not preserved in the larger Hilbert space with varying number of electrons. It is only preserved in the half-filled sector, acting according to:
	\begin{align}
	c_{m,\uparrow} \mapsto c^\dagger_{m,\downarrow},\quad c_{m,\downarrow} \mapsto - c^\dagger_{m,\uparrow},\cr
	c^\dagger_{m,\uparrow}\mapsto c_{m,\downarrow},\quad c^\dagger_{m,\downarrow}\mapsto -c_{m,\uparrow},
	\end{align}
	and on the empty fuzzy orbital $|0\rangle_m$, it acts as $|0\rangle_m \mapsto c^\dagger_{m,\downarrow}c^\dagger_{m,\uparrow}|0\rangle$. As \cite{Zhu:2022gjc} argue, this symmetry corresponds to the spacetime parity $P$ in the Ising CFT. Note that this symmetry flips the sign of the $SU(2)$ eigenvalue $L_z$.
\end{enumerate}

By studying the behavior of the Ising order parameter,
\begin{equation}
M = \sum_m \frac12 (c^\dagger_{m,\uparrow}c_{m,\uparrow} - c^\dagger_{m,\downarrow}c_{m,\downarrow}),
\end{equation}
the authors of \cite{Zhu:2022gjc} identified a line of critical points on the $(V_0, h)$ plane, which supports the Ising CFT. Let us quickly remind ourselves how this is done. One has to numerically determine the ground state of the Hamiltonian \eqref{Ham}, and compute the vacuum expectation value $\langle M^2\rangle$ of the Ising order parameter, as well as the binder cumulant $U_4 = \frac32 \left(1- \frac{\langle M^4\rangle}{3 \langle M^2\rangle^2}\right)$. By plotting these as functions of $(V_0, h)$ for varying system sizes (i.e., for different $N$, and including the appropriate scaling factor $N^{-2+\Delta_\sigma}$ for $\langle M^2\rangle$), one identifies the critical line as the one where the plots intersect. Then \cite{Zhu:2022gjc} pick a particular point $V_0=4.75$, $h_c=3.16$ on the critical line, which seems to minimize the finite-size effects. Numerical diagonalization at this point allows to approximately solve the 3d Ising model to, in principle, arbitrary precision, as long as we the number $N$ of electrons is high enough. In the rest of this paper, we will use the same values of $V_0$ and $h_c$ in the bulk, making sure that the bulk is at criticality. We will vary the boundary interaction and the boundary magnetic field as necessary.

In making these steps, \cite{Zhu:2022gjc} adjust Hamiltonian to the point where the finite size effects are almost negligible, and conformal spectrum emerges. In terms of some effective field theory (EFT) description, which (with infinitely many irrelevant interactions) is supposed to be valid even at finite $N$, this corresponds to adjusting the irrelevant couplings to the critical point. One notes, for example, that the density-density interaction $n^0(\Omega_a)n^0(\Omega_b)$, where $n^0 = n_\uparrow + n_\downarrow$, is one such irrelevant perturbation, which is already properly accounted for in the formula \eqref{Ham}.

However, the analysis based on the order parameter vev mentioned above does not guarantee the conformal spectrum of $H$. Even if $H$ exhibits such a spectrum, one can replace:
\begin{equation}
H \mapsto H + a L^2,
\end{equation}
where $L^2$ is the Casimir for spatial rotations. This will shift the spectrum, which will naively appear non-conformal. At the same time, it will not affect the ground state (and other eigenstates), and the vacuum expectation values $\langle M^2\rangle$ and $\langle M^4\rangle$ will remain the same. Therefore, with such a shifted Hamiltonian, we would still be able to find the same line of critical points, yet we would not see the conformal energy spectrum. This shows that if we want to recover the conformal spectrum, we might have to shift the Hamiltonian by the available conserved charges (without breaking any symmetries: e.g., on the full sphere, we can shift $H$ by $L^2$, but not, say, by $L_z$, as that would break both the $SU(2)$ and $P$). In the case of the full fuzzy sphere Hamiltonian \eqref{Ham}, the authors of \cite{Zhu:2022gjc} have already done this job for us. In other cases, especially in the hemisphere analysis below, we have to keep this possibility in mind, and potentially shift $H$ by the conserved charges in order to reduce the finite size effects and get a more accurate conformal spectrum.

\section{Boundary CFT from fuzzy hemisphere}\label{sec:fuzzy-BCFT}
The usual radial quantization technology applied to a theory on the half-space $\R^2\times \R_>$ leads to the hemisphere geometry:
\begin{equation}
HS^2 \times \R,
\end{equation}
where $HS^2$ is the hemisphere, with the appropriate boundary conditions. In this case, the spectrum of Hamiltonian on $HS^2$ also has the form $E^b_n = E^b_0 + \alpha \Delta^b_n$, with $E_0^b$ the Casimir energy in the presence of the boundary, and $\Delta_n^b$ the scaling dimensions of the boundary operators. Identification of $\Delta_n^b$, and the related data of bulk-boundary and boundary-boundary OPE coefficients, is of great interest, as the boundary criticality is ubiquitous.

An obvious extension of the techniques reviewed in Section \ref{sec:fuzzy-review} to the case of boundary CFT (BCFT) requires replacing $HS^2$ with the fuzzy hemisphere. Or does it not? In fact, there are two ways to go about it:
\begin{enumerate}
	\item Work with the full sphere, and impose constraints on the degrees of freedom supported on, say, the upper hemisphere, such that only the lower hemisphere achieves criticality, and the equator flows to the conformal boundary.
	\item Cut the sphere in half and actually work with the fuzzy hemisphere.
\end{enumerate}
Below, we will explore both options. As we will see, the first approach allows to preserve more symmetry, in particular the $P$ parity. The second approach explicitly breaks $P$, but we of course expect to regain it in the IR limit. Surprisingly, in our analysis we sometimes achieve better precision with the second method, which suggests that there is still room for improvement as far as utilizing symmetries and reducing the finite-size effects goes.

\subsection{Ising BCFT: expectations}
It is known that the 3d Ising CFT admits three types of conformal boundary conditions, depending on whether the 2d boundary remains ordered or disordered while the 3d bulk passes through the order/disorder phase transition. In the conventional description \cite{Cardy:1996xt,Diehl:1986,Diehl:1998mh,Deng:2005dh,Liendo:2012hy,Gliozzi:2015qsa}, one considers the classical statistical 3d Ising model at non-zero temperature, which passes through the classical critical point at $T=T_c$. The strength of nearest neighbor spin-spin interaction in the bulk is denoted by $J$. In the presence of boundary, one may consider the spin-spin interaction along the boundary surface $J_s\neq J$ different from the bulk one:
\begin{equation}
H = -J\sum_{\text{bulk} \langle ij\rangle} \sigma_i\sigma_j - J_s \sum_{\text{boundary} \langle ij\rangle} \sigma_i \sigma_j.
\end{equation}
Then by varying the surface interaction strength $J_s$, one interpolates between the three classes of boundary conditions:
\begin{figure}[h]
	\centering
	\includegraphics[scale=1]{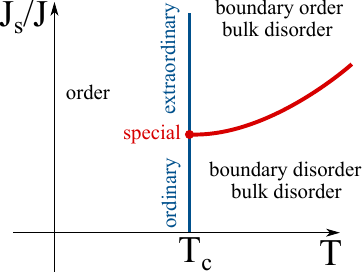}
	\caption{\label{fig:phase-class}Bulk/boundary phase diagram of the classical 3d Ising model.}
\end{figure}\\
 
When $J_s/J$ is large enough, the boundary orders before the bulk. This means that at the boundary, the $\Z_2$ symmetry is spontaneously broken, and for such parameters, the boundary flows to the so-called ``extraordinary'' boundary conditions \cite{Metlitski:2020cqy,Padayasi:2021sik,Toldin:2021kun,Trepanier:2023tvb}. Since $\Z_2$ is spontaneously broken at low energies, the theory decomposes into two sectors, with all boundary spins pointing up in one sector, and down -- in the other. Such boundary conditions have no relevant boundary deformations \cite{Gliozzi:2015qsa}. A sector with explicitly broken $\Z_2$ at the boundary and spins pointing up or down is known as the normal boundary criticality, so
\begin{equation}
\text{Extraordinary} = \text{normal} \oplus \overline{\text{normal}}.
\end{equation}
When $J_s/J$ is small enough (for example, we could turn $J_s$ off completely), the boundary remains disordered, and this engineers the ``ordinary'' boundary conditions. These are the $\Z_2$-preserving boundary conditions, which in the mean field theory description are realized by the Dirichlet boundary conditions on the bulk order parameter $\sigma$, that is $\sigma\big|=0$ \cite{Liendo:2012hy,Gliozzi:2015qsa}. The only boundary relevant operator corresponds to $\partial_\perp \sigma$ evaluated along the boundary.

Finally, between the ordinary and extraordinary regimes, at the critical value of $J_s$, sits the special boundary transition. In the mean field theory description, it is realized by the Neumann boundary conditions on $\sigma$, and supports two relevant operators, $\sigma$ and $\sigma^2$. In the BCFT, we will refer to them as $\hat\sigma$ and $:\hat\sigma^2:$. The classical bulk/boundary phase diagram on Figure \ref{fig:phase-class} clearly shows the three classes. Note that the boundary order parameter $\sigma_b$ (the boundary-averaged Ising spin) can be used to distinguish the three universality classes: $\langle \sigma_b^2\rangle=\frac14$ in the extraordinary, $\langle \sigma_b^2\rangle=0$ in the ordinary class, and $\langle\sigma_b^2 \rangle$ takes some nontrivial value in the special case. Thus we expect $\langle \sigma_b^2\rangle$, as a function of $J_s/J$, to interpolate between  $0$ and $\frac14$ as we change $J_s/J$ from zero to infinity.

In this paper, however, we follow \cite{Zhu:2022gjc} and realize the 3d Ising CFT as the quantum critical point of the $(2+1)d$ Ising-like system on the fuzzy sphere in transverse magnetic field $h$, at zero temperature. We expect to have the same qualitative picture of boundary criticality, with three universality classes, only the precise parameters are different. As mentioned before, the bulk criticality corresponds to a line in the $(V_0, h)$ parameter space. To navigate the classes of boundary crticality, we could either use the boundary transverse magnetic field $h_b$, or the boundary spin-spin interaction strength $v$ (which can be realized in a number of ways in the Hamiltonian, see below). We could then consider similar phase diagrams for the quantum criticality. For example, in the $(h, h_b)$ plane, we expect:
\begin{figure}[h]
	\centering
	\includegraphics[scale=1]{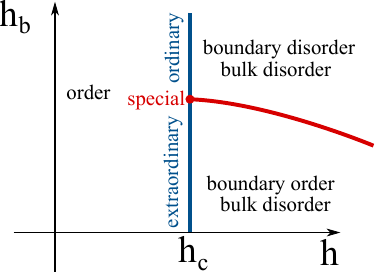}
	\caption{\label{fig:phase-quant}Phase diagram of the quantum critical Ising model, as a function of the bulk transverse magnetic field $h$ and the boundary transverse magnetic field $h_b$.}
\end{figure}\\
Similarly, we could consider the $\left(\frac1{V_0}, v_b \right)$ slice of the phase diagram, where $v_b \equiv v$ is the boundary spin-spin interaction strength, and it is more convenient to plot $\frac1{V_0}$ rather than $V_0$:
\begin{figure}[h]
	\centering
	\includegraphics[scale=1]{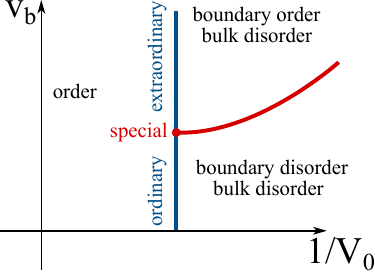}
	\caption{\label{fig:phase-quant1}Phase diagram of the quantum critical 3d Ising model, as a function of the bulk inverse interaction $V_0^{-1}$ and the boundary interaction $v_b$.}
\end{figure}

We will thus put $N/2$ electrons on the fuzzy hemisphere, adjust the bulk parameters $(V_0, h)$ to the critical point as in \cite{Zhu:2022gjc}, and use the boundary couplings $(h_b , v)$ (or just $h_b$) to drive the system into one of the three boundary classes. The continuum BCFT emerges in the $N\to\infty$ limit, in which the precise interactions do not matter due to universality (except for the special boundary conditions, which require fine-tuning). Alternatively, one may access the continuum BCFT at finite $N$ by fine-tuning a large number of irrelevant couplings in the Hamiltonian. This is of course the standard situation in effective field theory, as $N$ plays the role of UV cut-off. At finite $N$, the Hamiltonian is a finite square matrix, and we can adjust this matrix to have any spectrum we like (up to the energy cut-off). In particular, it can match the conformal spectrum. This requires fine tuning $O(N^2)$ many couplings, and such a theory looses its predictive power.

In practice, we need a sweet spot: We cannot make $N$ too large due to the computer limitations, and we cannot fine tune too many couplings for the same reason. For example, in the bulk, \cite{Zhu:2022gjc} choose to tune only two couplings, $V_0$ and $h$, while going up to $N=16$ (or $N=24$ via DMRG). We face the same dilemma in our study of the boundary criticality, and make the analogous choice of two boundary couplings to tune -- the boundary magnetic field $h_b$ and the boundary spin-spin coupling $v$. The magnetic field $h_b$ is rather easy to include. For even $N$, the orbital number $m=-\frac12$ is closest to the equator, and it couples to $h_b$ through:
\begin{equation}
\label{hb_def}
-h_b (c_{-1/2, \uparrow}^\dagger c_{-1/2, \downarrow} + c_{-1/2,\downarrow}^\dagger c_{-1/2,\uparrow}).
\end{equation}
We could also imagine the alternative ways of defining the $h_b$ coupling, such as evaluating contributions of all the orbitals to $n^x$ and integrating it precisely along the equator. While we do not claim that \eqref{hb_def} gives the best precision, it is certainly the simplest way to define the $h_b$ coupling, and it respects all the symmetries, so we stick to such a definition.

As for the spin-spin coupling $v$, there are several (inequivalent) ways to include it, without an obviously preferred one:
\begin{enumerate}
	\item We can simply add the following term coupling to the boundary orbital:
	\begin{equation}
	\label{v_1}
	v\cdot c^\dagger_{-1/2, \uparrow} c_{-1/2, \uparrow} c^\dagger_{-1/2, \downarrow} c_{-1/2, \downarrow}.
	\end{equation}
	This term is like $n_\downarrow n_\uparrow$ integrated along the boundary, except it would include contributions from all the orbitals, while in \eqref{v_1} we only keep the orbital $m=-\frac12$. The advantage of this term is that it is very easy to implement. This term preserves $\Z_2$ but is inconsistent with the $P$ symmetry.
	\item We can write a slightly different boundary interaction, which is like $n^z n^z$ integrated along the boundary:
	\begin{equation}
	\label{v_2}
	v\cdot (c^\dagger_{-1/2, \uparrow}c_{-1/2, \uparrow} - c^\dagger_{-1/2, \downarrow} c_{-1/2, \downarrow})^2.
	\end{equation}
	This term differs from \eqref{v_1} by the number operator $c^\dagger_{-1/2,\sigma}c_{-1/2,\sigma}$.
	The term \eqref{v_2} clearly favors the vacuum to involve the $\Z_2$-preserving states $|0\rangle$ and $c^\dagger_{-1/2, \uparrow} c^\dagger_{-1/2, \downarrow}|0\rangle$ for $v\gg 0$, and the $\Z_2$-breaking ones $c^\dagger_{-1/2,\uparrow}|0\rangle$ and $c^\dagger_{-1/2,\downarrow}|0\rangle$ for $v\ll 0$. In other words, we expect the $\Z_2$-preserving (``ordinary'') boundary phase for large positive $v$ and the $\Z_2$-breaking (``extraordinary'') boundary phase for large negative $v$, with the special transition sitting somewhere in between. Also, this interaction is compatible with both $\Z_2$ and $P$ symmetries.
	\item A slight variation of the coupling \eqref{v_1} is to shift the parameter $V_0 \mapsto V_0 + v$ (appearing in the bulk Hamiltonian \eqref{HamExpl},\eqref{V-coeffs}), but only in the terms describing the boundary spin-spin interaction. These are the terms containing the following combination:
	$$
	c^\dagger_{-1/2, \uparrow}c_{-1/2, \uparrow}c^\dagger_{-1/2, \downarrow}c_{-1/2, \downarrow}.
	$$
	Extracting such terms from the Hamiltonian and performing the shift, we isolate the boundary interaction as:
	\begin{equation}
	\label{v_1-variant}
	v\cdot 2(2N-1)\frac{2^{3N-2}\Gamma(N-1)\Gamma(\frac12(N+1))}{\Gamma(\frac12(N+2))\Gamma(2N)} c^\dagger_{-1/2, \uparrow} c_{-1/2, \uparrow} c^\dagger_{-1/2, \downarrow} c_{-1/2, \downarrow},
	\end{equation}
	which is simply \eqref{v_1} with a nontrivial $N$-dependent rescaling of $v$.
	\item We can also attempt to write the interaction literally given by $n^z n^z$ or $n_{\uparrow}n_{\downarrow}$ integrated along the boundary $\theta=\frac{\pi}{2}$:
	\begin{equation}
	\label{v_3}
	\int n^z(\varphi)n^z(\varphi) \dd\varphi =  \sum_{\substack{m_1+m_2\\= m_3+m_4}} W_{m_1}W_{m_2}W_{m_3}W_{m_4}(c^\dagger_{m_1,\uparrow} c_{m_4,\uparrow}-c^\dagger_{m_1,\downarrow} c_{m_4,\downarrow})(c^\dagger_{m_2,\uparrow} c_{m_3,\uparrow}-c^\dagger_{m_2,\downarrow} c_{m_3,\downarrow})
	\end{equation}
	\begin{equation}
	\label{v_4}
	\int n_\uparrow(\varphi)n_\downarrow(\varphi) \dd\varphi = \sum_{\substack{m_1+m_2\\= m_3+m_4}} W_{m_1}W_{m_2}W_{m_3}W_{m_4}c^\dagger_{m_1,\uparrow} c_{m_4,\uparrow}c^\dagger_{m_2,\downarrow}c_{m_3,\downarrow},
	\end{equation}
	where
	\begin{equation}
	W_m = \frac1{2^s}\sqrt{\frac{(2s+1)!}{4\pi(s+m)!(s-m)!}} = \frac1{2^s} N_m.
	\end{equation}
\end{enumerate}
We have experimented numerically with these options, and found that some of them give better precision, which we indicate below.

\subsection{Full sphere with constraints}\label{sec:frozen}
\paragraph{$\Z_2$-breaking boundary.} If we want to study the $\Z_2$-breaking, that is extraordinary, boundary conditions, we can limit ourselves to the case of normal boundary conditions (which describe one of the two superselection sectors in the extraordinary case). For that, we force each orbital in the upper hemisphere to contain precisely one (iso)spin, all of them pointing up. Such configurations form a $N\choose N/2$-dimensional subspace in the full Hilbert space on the sphere. This setup is illustrated in the Figure \ref{fig:half-up}, where the frozen spins on the upper hemisphere are shown in blue, and the spins on the lower hemisphere are dynamical:
\begin{figure}[h]
	\centering
	\includegraphics[scale=1]{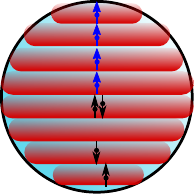}
	\caption{\label{fig:half-up} Spins in the upper hemisphere are frozen and pointing up. This mimics the normal $\Z_2$-breaking boundary conditions for the dynamical spins in the lower hemisphere.}
\end{figure}

This subspace is obviously not invariant under the Ising $\Z_2$, however, it is preserved by the $P$ parity, as the $|\uparrow\rangle_m$ state of $m$-th orbital is $P$-invariant. We can denote this subspace as $\cH_\uparrow$, and define a projector:
\begin{equation}
\Pi_\uparrow: \cH \to \cH_\uparrow.
\end{equation}
The Hamiltonian $H$ of the full sphere \eqref{Ham} is projected,
\begin{equation}
H_{\rm hemi} = \Pi_\uparrow H,
\end{equation}
to define the hemisphere system. In addition to $H_{\rm hemi}$, we may include the boundary magnetic field \eqref{hb_def} at the boundary, as well as the modified spin-spin interaction \eqref{v_2} (we cannot use \eqref{v_1} as it breaks the particle-hole symmetry $P$, and we do not use \eqref{v_3} as it makes numerics significantly slower). We will present the numerical results in Section \ref{sec:z2-break}.

Finally, we are also allowed to shift the Hamiltonian by an arbitrary function of conserved charges without breaking any symmetries. The symmetries we have in this case are the parity/particle-hole symmetry $P$ and the $U(1)$ rotations of the hemisphere, generated by $L_z$. We are allowed to shift the Hamiltonian by an arbitrary function of $L_z^2$:
\begin{equation}
\label{J2-freedom}
H_{\rm hemi} \mapsto H_{\rm hemi} + f(L_z^2).
\end{equation}
The reason is that $P$ flips the sign of $L_z$, so we can only add a function of $L_z^2$, or else we break the $P$ symmetry. This symmetry implies that the spectrum contains scaling operators in pairs, with the same conformal dimension but opposite $L_z$ eigenvalue for the operators in a pair.\footnote{Recall that on the full sphere, we have the $SU(2)$ multiplets of operators instead, where $L_z$ runs from $-j$ to $j$.} We are not allowed to shift $H_{\rm hemi}$ by $P$, as this would break $L_z$. We will use the freedom to perform shifts \eqref{J2-freedom} in Section \ref{sec:z2-break}: Indeed, the Hamiltonian $H_{\rm hemi} = \Pi_\uparrow H$ might differ from the conformal Hamiltonian on the hemisphere precisely by such shifts (as well as irrelevant interactions that we ignore). Such shifts correspond to the possibility of gravitational background counterterms localized at the boundary, which modify the stress-energy tensor and hence the Hamiltonian.

\paragraph{$\Z_2$-invariant boundary.} We can apply the same technique of freezing the upper hemisphere to study the $\Z_2$-preserving boundary conditions, particularly the ordinary class. For that purpose, we fix again the upper hemisphere to contain precisely one spin per orbital, but this time forcing them all to be in the $\Z_2$ even state, which for the $m$-th orbital is:
\begin{equation}
|\rightarrow\rangle_m = \frac1{\sqrt{2}}\left(|\uparrow\rangle_m + |\downarrow\rangle_m\right).
\end{equation}
This setup is illustrated in the Figure \ref{fig:half-right}:
\begin{figure}[h]
	\centering
	\includegraphics[scale=1]{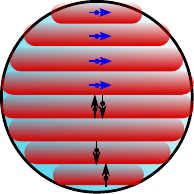}
	\caption{\label{fig:half-right} Each spin in the upper hemisphere is frozen to the $\Z_2$-even state.}
\end{figure}\\
Now we have that each state $|\rightarrow\rangle_m$ is both $\Z_2$ and $P$-invariant. The configurations with the upper hemisphere frozen to the $\Z_2$-even and $P$-invariant state again form an ${N\choose N/2}$-dimensional subspace of the full sphere Hilbert space $\cH$. Denoting it as $\cH_\rightarrow$, we define a projector:
\begin{equation}
\Pi_\rightarrow: \cH \to \cH_\rightarrow.
\end{equation}
The Hamiltonian for the $\Z_2$-even hemisphere system now is:
\begin{equation}
H_{\rm hemi} = \Pi_\rightarrow H.
\end{equation}
Again, we may additionally include the boundary magnetic field \eqref{hb_def}, and modify the boundary spin-spin interaction by adding \eqref{v_2}, \eqref{v_3} or \eqref{v_4}, which we will explore later. Since now the conserved charges include $L_z$ generating the $U(1)$, $P$, and the $\Z_2$ parity, we have more freedom in shifting $H_{\rm hemi}$ by the conserved charges. Namely, the allowed shifts are:
\begin{equation}
H_{\rm hemi} \mapsto H_{\rm hemi} + f_{\pm}(L_z^2),
\end{equation}
where the subscript in $f_{\pm}(L_z^2)$ corresponds to the $\Z_2$-even and $\Z_2$-odd subspaces. Denoting the $\Z_2$ charge as $Z$, which obeys $Z^2=1$, we can equivalently write this shift as:
\begin{equation}
\label{J2pm-freedom}
H_{\rm hemi} \mapsto H_{\rm hemi} + \frac12 (1+Z) f_+(L_z^2) + \frac12 (1-Z) f_-(L_z^2).
\end{equation}
A priori, there is no reason to expect that the Hamiltonian $H_{\rm hemi} = \Pi_\rightarrow H$ gives the correct conformal spectrum. Like in the $\Z_2$-breaking case, a shift \eqref{J2pm-freedom} might be necessary, and will be used in Section \ref{sec:z2-preserve} to improve the precision. The scaling operators will again come in $P$-multiplets, i.e., pairs of equal scaling dimensions and opposite $L_z$ charge that are swapped by $P$. The scalar operators (those with $L_z=0$) thus can be separated into the $P$-even and $P$-odd ones. The whole spectrum is also subdivided into the $\Z_2$-even and $\Z_2$-odd parts.

\subsection{Actual hemisphere}
We can also work with the actual fuzzy hemisphere, rather than the full sphere, half of which is frozen. Equivalently, this can be phrased as keeping the Northern hemisphere empty, with $N/2$ electrons living on the Southern hemisphere. Such setup is illustrated on Figure \ref{fig:true-hemi}:
\begin{figure}[h]
	\centering
	\includegraphics[scale=1]{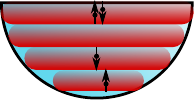}
	\caption{\label{fig:true-hemi} Hemisphere with $N/2$ fuzzy orbitals.}
\end{figure}\\
One major drawback of this setup is that it breaks the particle-hole symmetry $P$. We can use the same Hamiltonian \eqref{Ham} of the full sphere restricted to the Southern hemisphere to define the quantum system. Denote the Hilbert subspace with the Southern hemisphere half-filled (and empty Northern hemisphere) by $\cH_S$. There is a projector:
\begin{equation}
\Pi_S: \cH \to \cH_S,
\end{equation}
and the Hamiltonian is:
\begin{equation}
H_{\rm hemi} = \Pi_S H.
\end{equation}
Here we clearly see why such $H_{\rm hemi}$ breaks $P$: Not only $P$ maps $\cH_S$ to the different subspace, with the Northern hemisphere completely filled, but also $P$ is not a symmetry any more. Indeed, as was mentioned earlier, $P$ is only preserved in the half-filled subspace of $\cH$, whereas $\cH_S$ is a ``quarter-filled'' subspace of $\cH$.

Another problem is that the total angular momentum $L_z$ must be ``renormalized'' on the hemisphere. On the full sphere, when each orbital contains precisely one electron, the total $L_z$ eigenvalue is:
\begin{equation}
L_z = -s + (-s+1) + \dots + (s-1) + s = 0.
\end{equation}
In particular, when all spins are pointing down (or up), we get a $\Z_2$-degenerate vacuum state in the ferromagnetic phase. In the paramagnetic phase, all spins are in the $|\rightarrow\rangle$ state, and again the total angular moment is zero. Vacuum at the 3d Ising transition point, clearly, also has the angular momentum zero. Now, if we distribute all the electrons evenly over the hemisphere (one electron per orbital), we get the total $L_z$ eigenvalue:
\begin{equation}
L_z = -s + (-s+1) + \dots + \frac{-1}{2} = -\frac12 \left(s+\frac12\right)^2 = -\frac{N^2}{8}.
\end{equation}
The renormalized angular momentum on the hemisphere should then be defined as:
\begin{equation}
\widehat{L}_z = L_z + \frac{N^2}{8},
\end{equation}
which thus vanishes for the state with uniformly distributed electrons, and gives correct results in the numerical analysis. One good way to motivate such a renormalization is by comparing with the frozen hemisphere approach of Section \ref{sec:frozen}. From that point of view, the shift by $\frac{N^2}{8}$ is precisely the contribution of the frozen Northern hemisphere electrons to the total angular momentum.

Again, the Hamiltonian $H_{\rm hemi}$ can be modified by the boundary magnetic field \eqref{hb_def}, as well as any of the boundary spin-spin interactions \eqref{v_1}, \eqref{v_2}, \eqref{v_1-variant}, \eqref{v_3}, \eqref{v_4} (since $P$ is already broken). As before, it can also be shifted by the commuting conserved charges, without affecting the eigenstates (but changing the eigenvalues of $H_{\rm hemi}$). Because $P$ is no longer preserved, we are allowed to shift $H_{\rm hemi}$ by a function of $\widehat{L}_z$, not $\widehat{L}_z^2$, and again this can be done independently in the $\Z_2$-even and $\Z_2$-odd sectors:
\begin{equation}
H_{\rm hemi} \mapsto H_{\rm hemi} + f_{\pm}(\widehat{L}_z).
\end{equation}
Such shifts will play role later in Section \ref{sec:z2-preserve}, as they will be necessary to bring the spectrum of $H_{\rm hemi}$ closer to being conformal.

Despite obvious disadvantages of the hemisphere --- no $P$ symmetry, and the need to renormalize $L_z$ --- we will find that this approach often gives more precise numerics. This suggests that there must exist yet another, better approach, which would both preserve $P$ and give superior numerical precision.

\section{Numerical results}\label{sec:num}
Now let us describe in greater detail the numerics we do to study the ordinary and extraordinary (or normal) boundary conditions, as well as qualitative and preliminary quantitative results on the special boundary conditions. For the normal boundary conditions, we get the best results from the full sphere, whose Northern hemisphere is frozen with spins pointing up, as in Figure \ref{fig:half-up}. For the ordinary and special case, the true hemisphere of Figure \ref{fig:true-hemi} seems to give a slightly better precision, while the setup of Figure \ref{fig:half-right} allows to track the $P$ eigenvalues. We run our numerics in Python and in Mathematica, use the exact diagonalization only, and see indications of the BCFT spectrum already at relatively small $N$, up to $N=22$. Thus, clearly, there is a lot of room for improvement, such as optimizing the code and applying DMRG to push our results towards higher $N$.
\subsection{Sorcery to reduce the finite-size effects}\label{sec:sorcery}
At relatively small $N$, the finite-size effects become an issue, especially since on the hemisphere, we lose the protection of rotational symmetry $SU(2)$ of the full fuzzy sphere. Thus we need to find ways to reduce them. We use a number of tricks:
\begin{enumerate}
	\item Shifting Hamiltonian by the conserved charges to bring the spectrum closer to conformality. This one was explained earlier, and corresponds to the possibility of new background counterterms, that eventually contribute to the stress-energy tensor of the system. For each case in the sections below, we will explain exactly what shifts we perform.
	\item Scanning through the space of couplings to minimize some cost function that we take as a measure of conformality of the spectrum. In case we consider only one coupling (like the boundary magnetic field $h_b\in \R$), this is done by discretizing the range of values of this coupling, and identifying the value that minimizes the cost function. In case we consider two couplings, like $h_b$ and the boundary spin-spin coupling $v$, we perform the search as follows: Set the starting point $(h_b^{(0)}, v^{(0)})$, and alternate the minimization procedures, first minimizing in $h_b$, then in $v$, then in $h_b$ again etc. Each minimization consists of changing the value of that parameter by a given step $\pm\Delta h_b$ or $\pm\Delta v$, until we encounter the local minimum. The process runs until it converges. Then it is repeated for different initial points $(h_b^{(0)}, v^{(0)})$ to ensure that the minimum is global, discarding the local minima in the process.
	
	Adjusting the couplings brings us closer to the CFT point in the EFT description, and thus reduces the finite-size effects.
	\item By computing the scaling dimensions (Hamiltonian eigenvalues) of various operators at different values of $N$, we could extrapolate them as a series in the UV cutoff $\frac1{\sqrt{N}}$, finding (hopefully) a better approximation of the CFT data by plugging in a larger value of $N$. We were not able to get much improvement out of this technique yet, and will mostly provide the $N=22$ or $N=20$ answers as our best approximations.
\end{enumerate}

\subsection{$\Z_2$-breaking boundary}\label{sec:z2-break}
In principle, it should be possible to access the $\Z_2$-breaking boundary conditions on the true hemisphere of Figure \ref{fig:true-hemi} by, for example, tuning the boundary spin-spin interaction $v$ to a large enough value. In practice, the spontaneous $\Z_2$ symmetry breaking only occurs in the continuum limit. At finite $N$, it is hard to observe, and dynamically isolating the normal boundary conditions describing one of the $\Z_2$ vacua is even harder.

In this case, it is much easier to work on the full sphere, with the Northern hemisphere frozen as in Figure \ref{fig:half-up}. In addition to the basic Hamiltonian $H_{\rm hemi}= \Pi_\uparrow H$ defined earlier, we turn on the boundary magnetic field $h_b$ as in \eqref{hb_def}, and tune its value. For that we consider the orbital at $m=-\frac12$ as the ``boundary'', and define the boundary ``order'' parameter as measuring the spin on this orbital:
\begin{equation}
\label{b-order}
b = \frac12 (c^\dagger_{-1/2, \uparrow} c_{-1/2, \uparrow} - c^\dagger_{-1/2,\downarrow}c_{-1/2, \downarrow}).
\end{equation}
Of course, there is no true boundary order parameter in this case, since the $\Z_2$ symmetry is broken. We simply expect that $\langle b\rangle=\frac12$, as $b$ can be viewed as the boundary limit of the bulk order parameter. Thus it should have a vev $\frac12$, characterizing the regime with broken $\Z_2$. We compute the vev by literally taking the average of $b$ in the ground state of our Hamiltonian:
\begin{equation}
\langle b\rangle = \langle 0| b |0\rangle.
\end{equation}
We plot $\langle b\rangle$ as a function of $h_b$ for several values of $N$:
\begin{figure}[h]
	\centering
	\includegraphics[scale=0.4]{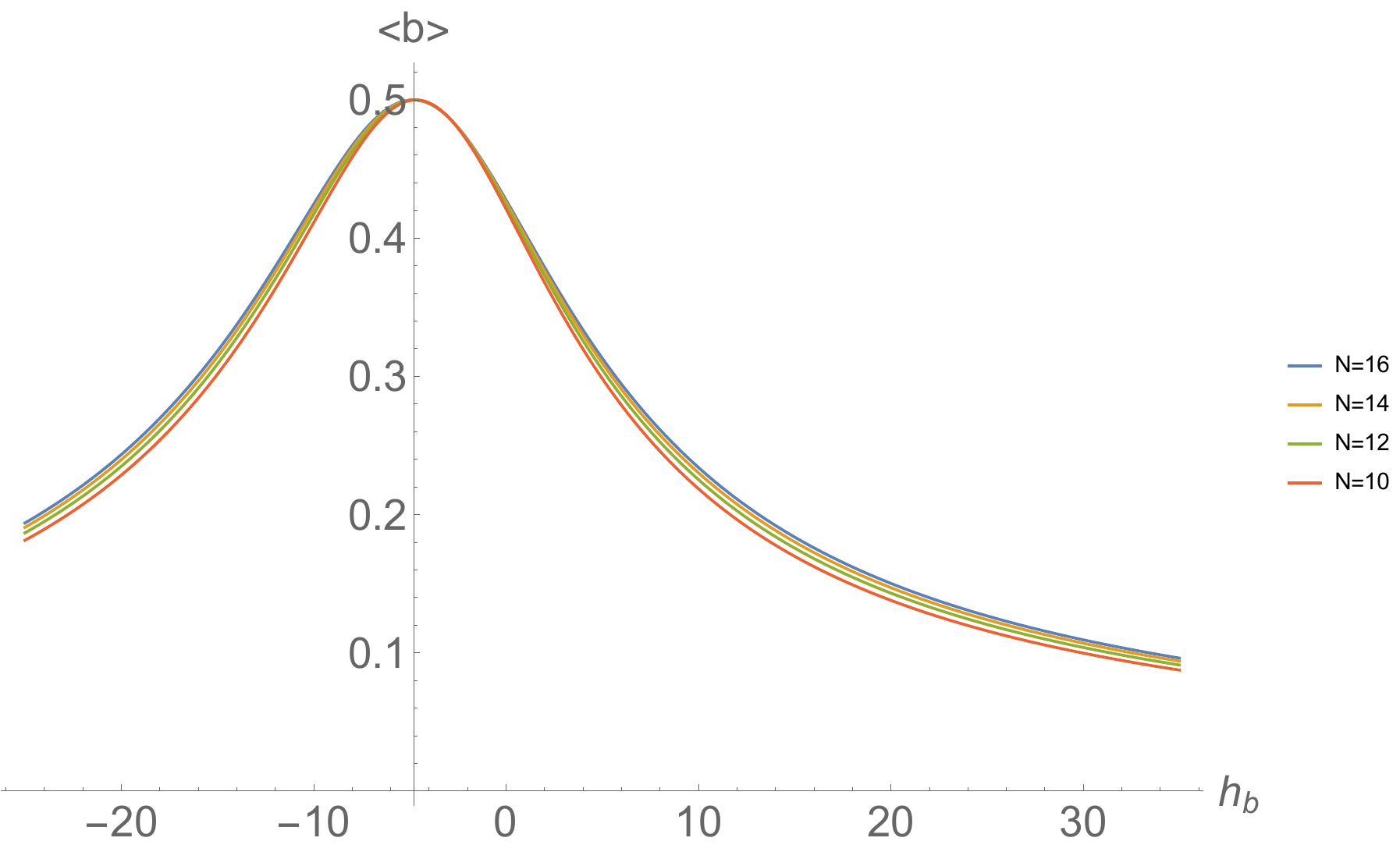}
	\caption{\label{fig:b-hb} We observe that the boundary vev $\langle b\rangle$ reaches its expected maximal value $\approx 0.5$ in the vicinity of $h_b=-4.8$.}
\end{figure}\\
We see that $\langle b\rangle$ reaches its maximal value close to $0.5$ somewhere around $h_b=-4.8$. Upon closer look, we find that the critical value of $h_b$, where $\langle b\rangle$ reaches its maximum, depends on $N$. We determine numerically the following table:
\begin{center}
	\begin{tabular}{ |c|c|c|c|c|c|c|c| } 
		\hline
		$N$ & 10 & 12 & 14 & 16 & 18 & 20 & 22 \\ 
		\hline
		$h_b$ & -4.6775 & -4.7470 & -4.7985 & -4.8386 & -4.8709 & -4.8978 & -4.9204 \\ 
		\hline
	\end{tabular}
\end{center}
where the critical value of $h_b$ has been determined with the $10^{-4}$ precision. Next we compute the spectrum of the Hamiltonian and estimate how close it is to the comformality. We use either the $N$-dependent critical value for $h_b$ given in the table, or the approximate value $h_b\approx -4.8$. In the final result, the difference is insignificant, however, the values given in the table lead to a slightly better precision. We plug in such values, compute the scaling dimensions up to $N=22$ (using the ED), and give the $N=22$ results in the table below as our best estimate. We provide the scaling dimensions of the first thirty operators at spins $L_z=0, L_z=\pm1, L_z=\pm2, L_z=\pm3$:

\begin{center}
	\begin{tabular}{ |c|c| } 
		\hline
		$L_z$ & Dimensions \\ 
		\hline
		0 & $\substack{\\ \\0,\ 3,\ 5.088,\ 5.858,\ 7.032,\ 7.819,\ 8.239,\ 8.498,\ 9.075,\ 9.328,\ 9.496,\ 9.860,\ 10.075,\ 10.139,\ 10.358,\ 10.379,\\ \boxed{10.433},\ 10.514,\ 10.605,\ 10.664,\ 10.684,\ 10.964,\ 11.202,\ 11.400,\ 11.631,\ \boxed{11.810},\ 11.927,\ 12.081,\ \boxed{12.097},\ 12.146}$ \\ 
		\hline
		1 & $\substack{4,\ 6.225,\ 6.788,\ 8.014,\ 8.783,\ 9.000,\ 9.085,\ 9.445,\ 9.721,\ 9.986,\ 10.168,\ 10.204,\ 10.302,\ 10.361,\ 10.451,\ 10.509,\\ 10.862,\ 10.983,\ 11.312,\ 11.533,\ 11.549,\ 11.622,\ 11.855,\ 11.974,\ 12.023,\ 12.159,\ 12.221,\ 12.237,\ 12.402,\ 12.535}$ \\ 
		\hline
		2 & $\substack{5,\ 7.148,\ 7.589,\ 7.816,\ 8.498,\ 9.265,\ 9.501,\ 9.682,\ 9.888,\ 9.908,\ 10.010,\ 10.036,\ 10.126,\ 10.182,\ 10.385,\ 10.662,\\ 10.719,\ 11.101,\ 11.227,\ 11.334,\ 11.656,\ 11.760,\ 11.904,\ 12.057,\ 12.077,\ 12.161,\ 12.186,\ 12.203,\ 12.255,\ 12.440}$ \\
		\hline
		3 & $\substack{6,\ 7.781,\ 8.400,\ 8.729,\ 8.815,\ 9.275,\ 9.517,\ 9.653,\ 9.744,\ 9.799,\ 9.985,\ 10.526,\ 10.681,\ 10.824,\ 10.987,\ 11.196,\\ 11.238,\ 11.287,\ 11.527,\ 11.580,\ 11.681,\ 11.692,\ 11.811,\ 11.887,\ 12.188,\ 12.298,\ 12.428,\ 12.545,\ 12.605, 12.736}$ \\
		\hline
	\end{tabular}
\end{center}
Here we used the prior knowledge \cite{Liendo:2012hy} that the lowest scalar operator at the normal/extraordinary boundary is the displacement operator $D\propto T_{\perp\perp}$, whose scaling dimension is exactly $3$. Thus, after shifting the energy spectrum such that the vacuum is at $0$, we rescale it to ensure that the first nontrivial scalar state has dimension $3$. As for the rest of dimensions, we used the freedom of adding $f(L_z^2)$ to the Hamiltonian: We shifted the conformal dimensions at spin $1$ to start at $\Delta=4$, corresponding to $\partial_i D$; at spin $2$ -- to start at $\Delta=5$, corresponding to $\partial_i\partial_j D$; and at spin $3$ -- to start at $\Delta=6$, corresponding to $\partial_i\partial_j\partial_k D$. Here $i, j, k$ denote the indices tangent to the boundary. Also note that a few scalar dimensions in the table are enclosed in boxes -- they indicate the $P$-odd scalar operators (i.e., pseudo-scalars).

We can easily identify states in the conformal multiplet of $D$ from the table:
\begin{align}
\label{D-data}
L_z = 0: \quad &\Delta = 3, \quad 5.088,\quad 7.032,\quad 9.075, &\text{ correspond to } D,\ \partial\bar\partial D,\ (\partial\bar\partial)^2 D,\ (\partial\bar\partial)^3 D\cr
L_z = 1: \quad &\Delta=4, \quad 6.225,\quad 8.014\qquad\qquad\qquad &\text{  must be } \partial D,\ \partial (\partial\bar\partial) D,\ \partial (\partial\bar\partial)^2 D\cr
L_z = 2: \quad &\Delta = 5, \quad 7.148\qquad\qquad\qquad\qquad\quad\ \ &\text{ must be } \partial^2 D,\ \partial^2(\partial\bar\partial) D\cr
L_z = 3: \quad &\Delta=6, \quad 7.781\qquad\qquad\qquad\qquad\quad\ \  &\text{ must be } \partial^3 D,\ \partial^3(\partial\bar\partial) D.
\end{align}
Here $\partial\bar\partial$ is the Laplace operator on the boundary, and $\partial$ is the partial derivative of spin $1$ along the boundary. As we move higher in dimension, the finite size effects make it harder to confidently identify which states are in the multiplet of $D$. A more advanced analysis resolving the finite size effects is required here. From the table of dimensions, without extra effort, we can also see part of the multiplet of the next scalar primary after $D$, -- call it $\cO$:
\begin{align}
L_z=0: \quad &\Delta=5.858,\quad 7.819,\quad 9.860\ &\text{ correspond to }  \cO,\ \partial\bar\partial \cO,\ (\partial\bar\partial)^2 \cO\cr
L_z=1: \quad &\Delta=6.788, \quad 8.783\ &\text{ must be } \partial \cO,\ \partial (\partial\bar\partial) \cO
\end{align}
while already at $L_z=2$, it is hard to identify $\partial^2 \cO$, as both dimensions $\Delta=7.589$ and $\Delta=7.816$ fit, -- more detailed analysis is required here.

We see from \eqref{D-data} that for the low-lying scalar operators, the dimensions of scalar descendants differ from the primary state by a multiple of $2$ with sufficiently good accuracy. This can be viewed as a signature of the conformal symmetry.

\subsection{$\Z_2$-preserving: ordinary}\label{sec:z2-preserve}
We can study the ordinary boundary conditions either by working on the hemisphere from Figure \ref{fig:true-hemi}, or on the full sphere with the Northern hemisphere frozen to the $\Z_2$-invariant state as in Figure \ref{fig:half-right}.

We have explored a number of possibilities, and here we focus on the full hemisphere with the frozen Northern hemisphere, and two boundary couplings. The latter include the boundary (or, rather, equator) magnetic field $h_b$ appearing through \eqref{hb_def}, and the boundary spin-spin interaction $v$ as in \eqref{v_2}. We fine-tune $(h_b, v)$ to make the spectrum look ``more conformal'' according to the criterion described below. For that, we use some prior expectations on the spectrum:
\begin{enumerate}
	\item In the scalar (spin $0$) sector, the spectrum starts from $0$ (vacuum), or course. The first non-trivial scalar is $\hat\sigma$, which describes the limit of the bulk order parameter (the latter vanishes at the ordinary boundary, and $\hat\sigma$ is the coefficient in front of the appropriate power of $r_\perp$). Its dimension $\Delta_{\hat\sigma}\approx 1.27$ is known in the bootstrap literature \cite{Gliozzi:2015qsa} (as well as from Monte Carlo \cite{hasenbusch2011thermodynamic} and perturbatively \cite{Diehl:1998mh}), to which we will compare later. The next scalar operator is supposed to be the displacement operator $D$, whose dimension is exactly $\Delta_D=3$, so we rescale the spectrum to match this condition. Finally, the fourth scalar is expected to be $\partial\bar\partial\hat\sigma$, -- the descendant of $\hat\sigma$. Its dimension is supposed to be precisely $\Delta_{\hat\sigma}+2$. This is one of the conditions we try to obey while scanning the range of possible values of $(h_b, v)$.
	\item In the vector (spin $1$) sector, the lowest state is suppsoed to be $\partial\hat\sigma$, followed by $\partial D$, and $\partial(\partial\bar\partial)\hat\sigma$. By using the freedom to shift the Hamiltonian by $f_\pm(L_z^2)$ (independently in the $\Z_2$-even and odd sectors), we can make sure that the dimension of $\partial\hat\sigma$ is precisely $\Delta_{\hat\sigma}+1$, while the dimension of $\partial D$ is precisely $4$. As for the dimension of $\partial(\partial\bar\partial)\hat\sigma$, it is supposed to be $\Delta_{\hat\sigma}+3$, however, we have no remaining freedom to achieve this by the shift. Thus, in scanning through the space of $(h_b, v)$ couplings, we are looking for the dimension of the third spin $1$ operator to be as close to $\Delta_{\hat\sigma}+3$ as possible.
\end{enumerate}
With these in mind, we scan over the space of couplings $(h_b, v)$, looking to obey the above two criteria. We do it for $N=12, 14, 16, 18, 20, 22$, and find the following optimal values:
\begin{center}
	\begin{tabular}{ |c|c|c|c|c|c|c| } 
		\hline
		$N$ & 12 & 14 & 16 & 18 & 20 & 22 \\ 
		\hline
		$h_b$ & -1.89972 & -2.689979 & -3.239104 & -3.667689 & -4.020210 & -4.319428 \\ 
		\hline
		$v$ & 11.2995 & 1.210587 & -0.257719 & -1.109985 & -1.723409 & -2.206263 \\ 
		\hline
	\end{tabular}
\end{center}
Note that the high precision of these values does \emph{not} reflect the actual precise values of these couplings at the critical point. These are simply the values that, within our numerical precision, guarantee that $\partial\bar\partial\hat\sigma$ sits $2$ units of energy above $\hat\sigma$ (with precision $10^{-6}$), and $\partial(\partial\bar\partial)\hat\sigma$ sits two units of energy above $\partial\hat\sigma$ (again with precision $10^{-6}$). The values we get are likely way less precise (in comparison to the exact critical point), due to the poorly controlled finite-size effects.

Our best estimate comes from the $N=22$ computation, and we provide the scaling dimensions below. Note that now all the operators also have the $\Z_2$ quantum number $\pm1$. For the $P$-odd scalar (pseudoscalar), we enclose the conformal dimension in a box. We only limit ourselves with the first $10$ dimensions in each sector:
\begin{center}
	\begin{tabular}{ |c|c|c| } 
		\hline
		$L_z$ & $\Z_2$ & Dimensions \\ 
		\hline
		0 & $+1$ & 0, 3, 4.89, 5.31, 6.28, 6.66, \boxed{6.84}, 7.17, 7.42, 7.51  \\ 
		\cline{2-3}
		  & $-1$ & 1.26, 3.26, 4.81, 5.07, 6.10, 6.86, 7.10,  7.35, 7.53, 7.59  \\
		\hline
		1 & $+1$ & 4, 5.57, 5.77, 6.27, 7.00, 7.09, 7.49, 7.76, 7.93, 8.05  \\ 
		\cline{2-3}
		& $-1$ & 2.26, 4.26, 5.64, 6.05, 6.67, 7.29, 7.46, 7.47, 7.49, 7.58  \\
		\hline
		2 & $+1$ & 5, 5.36, 6.76, 6.93, 7.14, 7.46, 7.88, 8.20, 8.51, 8.60  \\ 
		\cline{2-3}
		& $-1$ & 3.26, 4.91, 6.07, 6.67, 6.88, 6.95, 7.29, 7.33, 7.34, 7.41  \\
		\hline
		3 & $+1$ & 6, 6.12, 7.25, 7.49, 7.80, 8.00, 8.15, 8.28, 8.61, 8.75  \\ 
		\cline{2-3}
		& $-1$ & 4.26, 5.50, 6.63, 7.15, 7.21, 7.33, 7.34, 7.40, 7.50, 7.55  \\
		\hline
	\end{tabular}
\end{center}
We see that we obtain $\Delta_{\hat\sigma}\approx 1.26$, which agrees with the known value \cite{Gliozzi:2015qsa,hasenbusch2011thermodynamic,Diehl:1998mh}. The spectrum in this table, however, contains large finite-size effects and exhibits very poor conformality. We can recognize the following descendants of $\hat\sigma$:
\begin{align}
L_z=0,\quad \Delta=3.26, 4.81:\quad &\partial\bar\partial\hat\sigma,\quad (\partial\bar\partial)^2\hat\sigma,\cr
L_z=1,\quad \Delta=2.26, 4.26:\quad &\partial\hat\sigma,\quad \partial(\partial\bar\partial)\hat\sigma,\cr
L_z=2,\quad \Delta=3.26, 4.91:\quad &\partial^2\hat\sigma,\quad \partial^2(\partial\bar\partial)\sigma,\cr
L_z=3,\quad \Delta=4.26, 5.50:\quad &\partial^3\hat\sigma,\quad \partial^3(\partial\bar\partial)\hat\sigma
\end{align}
While for the states with dimensions $2.26=\Delta_{\hat\sigma}+1$, $3.26=\Delta_{\hat\sigma}+2$, and $4.26=\Delta_{\hat\sigma}+3$, the identification of operators is rather clear, how do we identify the rest? Say for $L_z=0$, the third $\Z_2$-odd state, with $\Delta=4.81$, was computed at $N=22$, while its predecessors at $N=14, 16, 18, 20$ have $\Delta=4.69, 4.74, 4.77, 4.79$. Thus the dimension of this state, evidently, grows with $N$. At the same time, the fourth $\Z_2$-odd scalar state at $N=14, 16, 18, 20, 22$ has $\Delta= 5.09, 5.08, 5.08, 5.08, 5.07$, which, evidently, slowly decreases with the growth of $N$. Thus the only reasonable conjecture is that the third $\Z_2$-odd scalar state reaches $\Delta_{\hat\sigma}+4\approx 5.26$ in the $N\to \infty$ limit. Similar reasoning applies to the second $\Z_2$-odd spin $2$ state of $\Delta=4.91$: This state at $N=14, 16, 18, 20, 22$ has monotonically increasing dimensions $\Delta=4.81, 4.84, 4.87, 4.89, 4.91$, whereas the third $\Z_2$-odd spin $2$ state has $\Delta=5.41, 5.64, 5.83, 5.97, 6.07$. The latter also grows with $N$, however, it is significantly larger than the expected $\Delta_{\hat\sigma}+4\approx 5.26$. Thus it only makes sense to assume that the second $\Z_2$-odd state is $\partial^2 (\partial\bar\partial)\hat\sigma$. The same exact argument works for the second $\Z_2$-odd state of spin $3$, whose dimension is $\Delta=5.50$ at $N=22$, it grows, and is expected to reach $\Delta_{\hat\sigma}+5\approx 6.26$ at $N\to\infty$.

Using similar arguments, we can also identify some states from the conformal multiplet of the displacement operator, which is the state at $L_z=0$ and $\Delta=3$. Of course the states at $\Delta=4, 5, 6$ correspond to $\partial D$, $\partial^2 D$ and $\partial^3 D$. The state at $L_z=0$ and $\Delta=4.89$ is in fact $\partial\bar\partial D$. It is hard to make further guesses due to the large finite-size effects.

\paragraph{Alternative realizations.} We saw that the full sphere with spins in the Northern hemisphere frozen, and couplings $(h_b, v)$ as in \eqref{hb_def}, \eqref{v_2}, provides a viable approach, albeit the precision is quite low. We explored various alternatives, which include using other couplings and working on the actual hemisphere. We would like to point out three alternative approaches that give decent results:
\begin{enumerate}
	\item Hemisphere, with the boundary magnetic field $h_b$ as in \eqref{hb_def}, and the boundary interaction $v$ as in \eqref{v_2}.
	\item Hemisphere, with the boundary magnetic field $h_b$ as in \eqref{hb_def}, and the boundary interaction $v$ as in \eqref{v_4}.
	\item Full sphere with the frozen upper hemisphere, again with the boundary magnetic field \eqref{hb_def}, but the boundary interaction is taken to be \eqref{v_3}.
\end{enumerate}
The first two approaches explicitly break the $P$ symmetry. Despite that, they tend to give better numerical results (spectrum that obeys conformality with higher precision) than the full sphere approach we used earlier. This suggests that the numerics can be further improved, both preserving $P$ and giving a higher precision. The third approach on the list above --- full sphere with the upper half frozen and with the $n^z n^z$ interaction along the boundary --- appears quite promising, as it gives great numerical results, while preserving the $P$ symmetry. Below we provide the first 10 scalar dimensions obtained using this approach at $N=20$:
\begin{center}
	\begin{tabular}{ |c|c|c|c|c|c|c|c|c|c|c| } 
		\hline
		$\Delta$ & 0 & 1.25 & 3 & 3.27 & 4.60 & 4.95 & 5.10 & 5.44 & 6.04 & 6.12 \\ 
		\hline
		$\Z_2$ & $+$ & $-$ & $+$ & $-$ & $-$ & $+$ & $-$ & $+$ & $-$ & $+$ \\ 
		\hline
		$P$ & $+$ & $+$ & $+$ & $+$ & $+$ & $+$ & $+$ & $+$ & $+$ & $+$ \\ 
		\hline
	\end{tabular}
\end{center}
where the $\Z_2$ and $P$ charges are also indicated. Here we point out that the descendant $\partial\bar\partial D$ of the displacement operator is clearly identified with the state of dimension $\Delta=4.95$. The difference from $3$ is $1.95$, which deviates from the expected difference of $2$ by only the $2.5\%$ error. We do not further investigate this case in this paper (the $n^z n^z$ interaction appears slightly more computationally heavy). However, it would be interesting to develop computational tools that would allow to efficiently identify the interactions giving the best approximation of the conformal point.

\subsection{$\Z_2$-preserving: special}\label{sec:special}
Finally, let us discuss the special boundary transition. We only have partial results in this case: While able to see characteristic features of the special transition, we could not obtain the reliable estimates of the scaling dimensions.

The special transition is a 2d Ising-like transition, living between the $\Z_2$-preserving (ordinary) and $\Z_2$-breaking (extraordinary) boundary phases. Therefore, it makes sense to characterize it in terms of the boundary order parameter $b$ introduced earlier in \eqref{b-order}, or rather the vacuum average of its square:
\begin{equation}
\langle b^2\rangle = \langle 0| b^2 |0\rangle,
\end{equation}
Just like in the bulk \cite{Zhu:2022gjc}, the scaling of this order parameter with $N$ can be used to characterize the \emph{special} BCFT point. Namely, we expect $\langle b^2 \rangle \propto N^{1-\Delta_{\hat\sigma}}$.\footnote{Compare this with $\langle M^2\rangle \propto N^{2-\Delta_\sigma}$ in \cite{Zhu:2022gjc}, where $M$ is the total Ising spin on the fuzzy sphere. Since $N$ plays the role of area on the fuzzy sphere, $M/N$ is spin per unit area, which is the order parameter $\sigma$ in the mean field theory, with dimension $\Delta_\sigma$ at the conformal point. Thus it is expected that $\langle\sigma^2\rangle \propto \sqrt{N}^{-2\Delta_\sigma}$, where $\sqrt{N}$ is the length parameter. In the BCFT, $b$ is the total boundary spin, while the boundary spin per unit length is the mean field $\hat\sigma$, implying $\langle b^2 \rangle \propto N^{1-\Delta_{\hat\sigma}}$.} Thus $\langle b^2 \rangle/ N^{1-\Delta_{\hat\sigma}}$ must be $N$-independent at the conformal point. We take the value $\Delta_{\hat\sigma}\approx 0.42$ from \cite{Liendo:2012hy}.

As in \cite{Zhu:2022gjc}, we could attempt to plot $\langle b^2 \rangle/ N^{1-\Delta_{\hat\sigma}}$, (or, alternatively, the boundary binder cumulant $U_4$,) as a function of couplings $(h_b, v)$, for different values of $N$. The plots are expected to intersect along the critical line, which is where the special boundary transition is located. This would allow to identify the critical values of $(h_b, v)$, and study the special boundary conditions. In practice, however, we were not able to complete this program yet. 

Here we present partial results. One promising attempt resulted from using the boundary spin-spin interaction given in \eqref{v_1-variant}, in addition to our usual boundary magnetic field $h_b$. The order parameter $\langle b^2 \rangle$ interpolates between $0$ for large negative $v$, and $1$ for positive $v$. Plotting $\langle b^2 \rangle/ N^{1-\Delta_{\hat\sigma}}$ as a function of $(h_b, v)$ for $N=12, 14, 16, 18, 20$, we find a promising pattern:
\begin{figure}[h]
	\centering
	\includegraphics[scale=0.4]{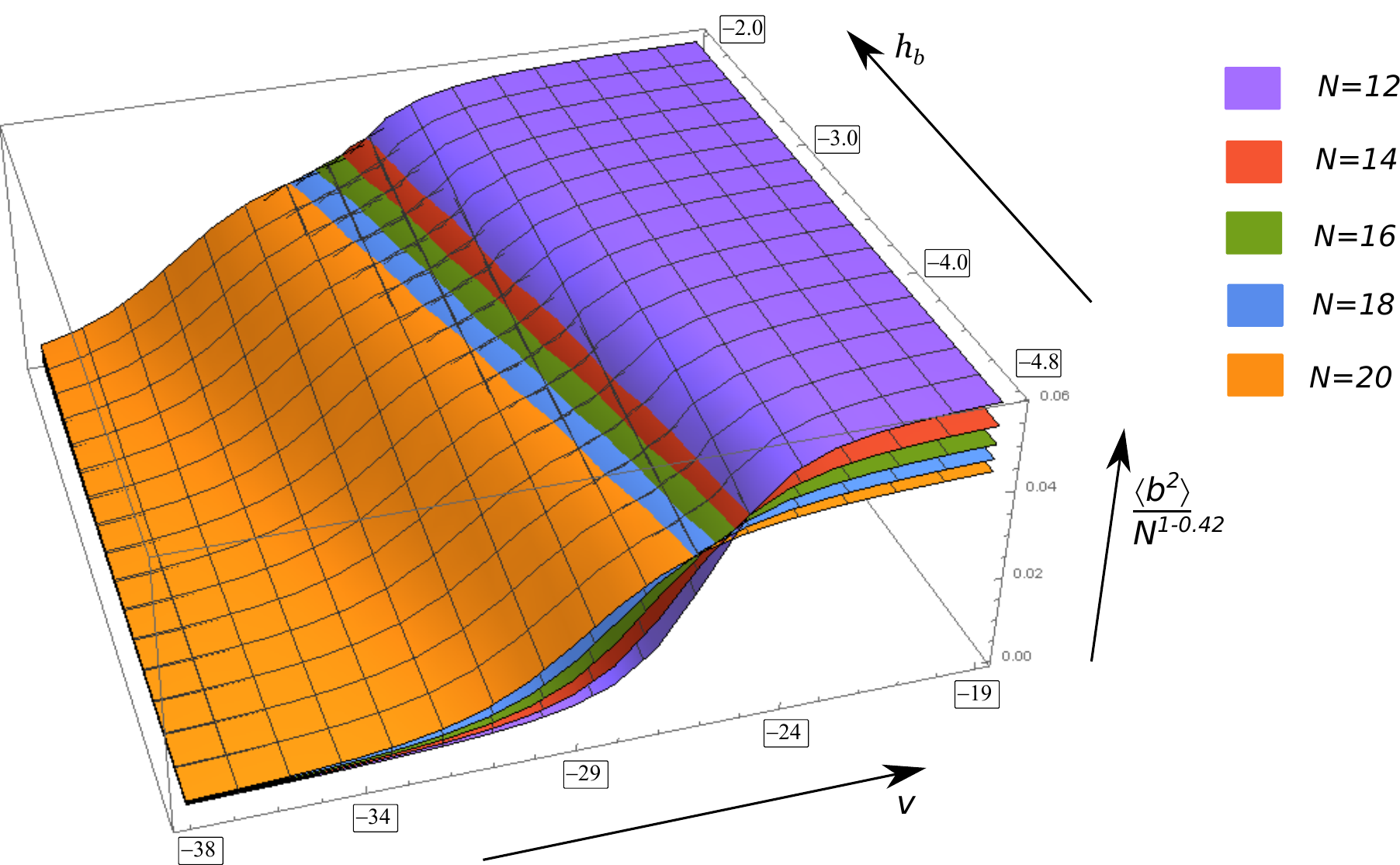}
	\caption{\label{fig:order-plot-special}Plots of $\langle b^2 \rangle/ N^{1-\Delta_{\hat\sigma}}$ as a function of $(h_b, v)$, for $N=12, 14, 16, 18, 20$.}
\end{figure}\\
All the plots approximately intersect along one line, indicating that the special transition might be located there. However, such an approximation is very crude. A closer look reveals that it is not really one line: It is a collection of several lines. The reason for this is that the critical values of $(h_b, v)$ receive $1/\sqrt{N}$ corrections, so at each $N$, the line of special boundary transition lies at a different location. A more complete analysis should take such finite size effects into account.

Nevertheless, we use our data to attempt the search for the special BCFT point in the vicinity of the intersection lines. We also use prior expectations on the spectrum. We know that the scalar spectrum should start with the $\Z_2$-odd operator $\hat\sigma$ of $\Delta_{\hat\sigma}\approx 0.42$, followed by the $\Z_2$-even $:\hat\sigma^2:$, $\Z_2$-odd $\partial\bar\partial \hat\sigma$, and the displacement operator $D$ of $\Delta_D=3$. In addition, according ot our previous discussions, we are allowed to shift energies independently in the $\Z_2$-even and $\Z_2$-odd sectors, for each spin. The $\Z_2$-even spin $0$ sector is fixed by the requirement that the vacuum has zero energy, whereas the rest may be shifted according to the conformal expectations.

We approximately identify a few points at $N=20$, and evaluate the spectrum. These include $(h_b, v) = (-1.9, -24)$, where the scalar spectrum is:
\begin{center}
	\begin{tabular}{ |c|c|c|c|c|c|c|c|c|c|c|c| } 
		\hline
		$\Delta$ & 0  & 0.42 & 2.20 & 2.44 & 3 & 3.46 & 4.28 & 4.30 & 4.48 & 4.90 & 4.97 \\ 
		\hline
		$\Z_2$ & $+$ & $-$ & $+$ & $-$ & $+$ & $-$ & $-$ & $+$ & $-$ & $-$ & $+$ \\ 
		\hline
	\end{tabular}
\end{center}
We also experimented with adding a second spin-spin interaction \eqref{v_4}, with the coefficient $v_1$ in front of it. For $(h_b, v, v_1)=(-2.6, -24, -3.563)$, we get:
\begin{center}
	\begin{tabular}{ |c|c|c|c|c|c|c|c|c|c|c|c| } 
		\hline
		$\Delta$ & 0  & 0.42 & 1.93 & 2.42 & 3 & 3.40 & 4.16 & 4.37 & 4.61 & 4.92 & 4.99 \\ 
		\hline
		$\Z_2$ & $+$ & $-$ & $+$ & $-$ & $+$ & $-$ & $+$ & $-$ & $-$ & $-$ & $+$ \\ 
		\hline
	\end{tabular}
\end{center}
For $(h_b, v, v_1)=(-2.6, -26, -2.771)$:
\begin{center}
	\begin{tabular}{ |c|c|c|c|c|c|c|c|c|c|c|c| } 
		\hline
		$\Delta$ & 0  & 0.42 & 1.57 & 2.42 & 3 & 3.03 & 3.78 & 4.32 & 4.45 & 4.64 & 4.72 \\ 
		\hline
		$\Z_2$ & $+$ & $-$ & $+$ & $-$ & $+$ & $-$ & $+$ & $-$ & $-$ & $+$ & $-$ \\ 
		\hline
	\end{tabular}
\end{center}

We see that all these spectra show features expected at the special transition: (1) the first non-trivial scalar is $\Z_2$-odd, corresponding to $\hat\sigma$; (2) it is followed by the $\Z_2$-even scalar that we loosely refer to as $:\hat\sigma^2:$, which is then followed by the $\Z_2$-odd $\partial \bar\partial \hat\sigma$; (3) the next state corresponds to the displacement operator $D$ at $\Delta=3$; (4) we can also clearly see candidates for $\partial\bar\partial :\hat\sigma^2:$ and for $\partial\bar\partial D$. However, the above numbers are clearly unreliable, since the dimension of the second scalar $\Delta_{:\hat\sigma^2:}$ takes a large range of values. The value $\Delta_{:\hat\sigma^2:}=2.2$ in the first table is clearly wrong, since $:\hat\sigma^2:$ is expected to be a relevant boundary deformation, hence $\Delta_{:\hat\sigma^2:}<2$. By choosing different values for $(h_b, v)$, we are able to tune $\Delta_{:\hat\sigma^2:}$ to any value between $1$ and $2$ (and, possibly, even below $1$). Thus, our analysis at this point is insufficient to give an estimate of $\Delta_{:\hat\sigma^2:}$.

Note that the bootstrap literature \cite{Liendo:2012hy,Gliozzi:2015qsa} also lacks a reliable estimate for the second relevant scalar at the special boundary. As for the Monte Carlo simulations \cite{Deng:2005dh}, their value for $\Delta_{\hat\sigma}$ appears to disagree with the bootstrap estimate $\Delta_{\hat\sigma}=0.42$ (which we take as more reliable, since the bootstrap provides rigorous numeric bounds). Therefore, to this date, the reliable estimate of $\Delta_{:\hat\sigma^2:}$ seems missing.

\section{Outlook}
The analysis of this paper shows that the fuzzy sphere techniques in 3d CFT can be extended to the case of BCFT, albeit at the cost of increased finite-size effects, as we inevitably break the rotational group $SO(3)$ down to $SO(2)$. We studied the example of 3d Ising BCFT in this paper, and were able to achieve decent numerical control over the two (out of three) classes of its conformal boundaries, known as the ``ordinary'' and ``extraordinary'' (or ``normal''). For the third, ``special'' boundary universality class, we were only able to see some of its qualitative features, however, the numerics seems inconclusive as far as the precise scaling dimensions of the boundary operators are concerned. Further studies are required to better pin down the properties of the special class, especially to eliminate the finite size effects, perhaps using the conformal perturbation theory as in \cite{Lao:2023zis}. Such studies would also help to properly estimate the error bars in our analyses (which we do not do in this paper).

We should note that such studies often involve searching through the spaces of couplings, attempting to make the spectrum ``more conformal'', which we also refer to as reducing the finite size effects. Including more irrelevant couplings in the Hamiltonian and optimizing over them, generally, would lead to better numerical results and more precise scaling dimensions. This does not only apply to the analysis of Ising BCFT, but also more generally to other 3d CFTs beyond the 3d Ising model. At the moment, only one other example has been explored in the literature \cite{Zhou:2023qfi,Chen:2024jxe}.

Indeed, if we want to study other 3d CFTs, we need to find the appropriate system of electrons on $\fs^2_N$, and tune their interactions to the quantum critical point, such that the spectrum becomes conformal. Performing such searches can be rather laborious, especially if we increase the number of flavors each electron has (in the Ising case, there are only two flavors: ``isospin up'' and ``isospin down''). Such a task should definitely be automated, and it is very tempting to propose that the machine learning could be of use. Training the network to recognize ``conformal features'' and subsequently scanning over the vast spaces of couplings, searching for fuzzy 3d CFTs, could be an interesting project to pursue. Another direction, that has already been proposed in the literature \cite{Zhu:2022gjc}, is to extend these techniques beyond three dimensions.
\subsection*{Acknowledgements}
I want to thank Yin-Chen He for helpful discussions on the subject. I am deeply grateful to Zheng Zhou and Yijian Zou for informing me about their results and agreeing to coordinate the submissions of our papers, as well as providing comments on this draft.

\bibliographystyle{utphys}
\bibliography{fuzzy-refs}

\providecommand{\href}[2]{#2}\begingroup\raggedright\begin{thebibliography}{10}

\bibitem{Belavin:1984vu}
A.~A. Belavin, A.~M. Polyakov, and A.~B. Zamolodchikov, ``{Infinite Conformal
  Symmetry in Two-Dimensional Quantum Field Theory},''
  \href{http://dx.doi.org/10.1016/0550-3213(84)90052-X}{{\em Nucl. Phys. B}
  {\bfseries 241} (1984) 333--380}.

\bibitem{Poland:2018epd}
D.~Poland, S.~Rychkov, and A.~Vichi, ``{The Conformal Bootstrap: Theory,
  Numerical Techniques, and Applications},''
  \href{http://dx.doi.org/10.1103/RevModPhys.91.015002}{{\em Rev. Mod. Phys.}
  {\bfseries 91} (2019) 015002},
  \href{http://arxiv.org/abs/1805.04405}{{\ttfamily arXiv:1805.04405
  [hep-th]}}.

\bibitem{Deng:2005dh}
Y.~J. Deng, H.~W.~J. Blote, and M.~P. Nightingale, ``{Surface and bulk
  transitions in three-dimensional O(n) models},''
  \href{http://dx.doi.org/10.1103/PhysRevE.72.016128}{{\em Phys. Rev. E}
  {\bfseries 72} (2005) 016128--016138},
  \href{http://arxiv.org/abs/cond-mat/0504173}{{\ttfamily
  arXiv:cond-mat/0504173}}.

\bibitem{Cosme:2015cxa}
C.~Cosme, J.~M. V.~P. Lopes, and J.~Penedones, ``{Conformal symmetry of the
  critical 3D Ising model inside a sphere},''
  \href{http://dx.doi.org/10.1007/JHEP08(2015)022}{{\em JHEP} {\bfseries 08}
  (2015) 022}, \href{http://arxiv.org/abs/1503.02011}{{\ttfamily
  arXiv:1503.02011 [hep-th]}}.

\bibitem{ferrenberg2018pushing}
A.~M. Ferrenberg, J.~Xu, and D.~P. Landau, ``{Pushing the limits of Monte Carlo
  simulations for the three-dimensional Ising model},'' {\em Physical Review E}
  {\bfseries 97} no.~4, (2018) 043301.

\bibitem{Meneses:2018xpu}
S.~a. Meneses, J.~a. Penedones, S.~Rychkov, J.~M. Viana Parente~Lopes, and
  P.~Yvernay, ``{A structural test for the conformal invariance of the critical
  3d Ising model},'' \href{http://dx.doi.org/10.1007/JHEP04(2019)115}{{\em
  JHEP} {\bfseries 04} (2019) 115},
  \href{http://arxiv.org/abs/1802.02319}{{\ttfamily arXiv:1802.02319
  [hep-th]}}.

\bibitem{Lao:2023zis}
B.-X. Lao and S.~Rychkov, ``{3D Ising CFT and exact diagonalization on
  icosahedron: The power of conformal perturbation theory},''
  \href{http://dx.doi.org/10.21468/SciPostPhys.15.6.243}{{\em SciPost Phys.}
  {\bfseries 15} no.~6, (2023) 243},
  \href{http://arxiv.org/abs/2307.02540}{{\ttfamily arXiv:2307.02540
  [hep-th]}}.

\bibitem{Zhu:2022gjc}
W.~Zhu, C.~Han, E.~Huffman, J.~S. Hofmann, and Y.-C. He, ``{Uncovering
  Conformal Symmetry in the 3D Ising Transition: State-Operator Correspondence
  from a Quantum Fuzzy Sphere Regularization},''
  \href{http://dx.doi.org/10.1103/PhysRevX.13.021009}{{\em Phys. Rev. X}
  {\bfseries 13} no.~2, (2023) 021009},
  \href{http://arxiv.org/abs/2210.13482}{{\ttfamily arXiv:2210.13482
  [cond-mat.stat-mech]}}.

\bibitem{Hu:2023xak}
L.~Hu, Y.-C. He, and W.~Zhu, ``{Operator Product Expansion Coefficients of the
  3D Ising Criticality via Quantum Fuzzy Spheres},''
  \href{http://dx.doi.org/10.1103/PhysRevLett.131.031601}{{\em Phys. Rev.
  Lett.} {\bfseries 131} no.~3, (2023) 031601},
  \href{http://arxiv.org/abs/2303.08844}{{\ttfamily arXiv:2303.08844
  [cond-mat.stat-mech]}}.

\bibitem{Han:2023yyb}
C.~Han, L.~Hu, W.~Zhu, and Y.-C. He, ``{Conformal four-point correlators of the
  three-dimensional Ising transition via the quantum fuzzy sphere},''
  \href{http://dx.doi.org/10.1103/PhysRevB.108.235123}{{\em Phys. Rev. B}
  {\bfseries 108} no.~23, (2023) 235123},
  \href{http://arxiv.org/abs/2306.04681}{{\ttfamily arXiv:2306.04681
  [cond-mat.stat-mech]}}.

\bibitem{Hofmann:2023llr}
J.~S. Hofmann, F.~Goth, W.~Zhu, Y.-C. He, and E.~Huffman, ``{Quantum Monte
  Carlo Simulation of the 3D Ising Transition on the Fuzzy Sphere},''
  \href{http://dx.doi.org/10.21468/SciPostPhysCore.7.2.028}{{\em SciPost Phys.
  Core} {\bfseries 7} (2024) 028},
  \href{http://arxiv.org/abs/2310.19880}{{\ttfamily arXiv:2310.19880
  [cond-mat.str-el]}}.

\bibitem{Rychkov:2016iqz}
S.~Rychkov, \href{http://dx.doi.org/10.1007/978-3-319-43626-5}{{\em {EPFL
  Lectures on Conformal Field Theory in D\ensuremath{>}= 3 Dimensions}}}.
\newblock SpringerBriefs in Physics. 1, 2016.
\newblock \href{http://arxiv.org/abs/1601.05000}{{\ttfamily arXiv:1601.05000
  [hep-th]}}.

\bibitem{MWeigel_2000}
M.~Weigel and W.~Janke, ``{Universal amplitude-exponent relation for the Ising
  model on sphere-like lattices},''
  \href{http://dx.doi.org/10.1209/epl/i2000-00377-0}{{\em Europhysics Letters}
  {\bfseries 51} no.~5, (Sep, 2000) 578}.

\bibitem{Deng:2002ea}
Y.~J. Deng and H.~W.~J. Bloete, ``{Conformal invariance of the Ising model in
  three-dimensions},''
  \href{http://dx.doi.org/10.1103/PhysRevLett.88.190602}{{\em Phys. Rev. Lett.}
  {\bfseries 88} (2002) 190602}.

\bibitem{Brower:2024otr}
R.~C. Brower and E.~K. Owen, ``{The Ising Model on $\mathbb S^2$},''
  \href{http://arxiv.org/abs/2407.00459}{{\ttfamily arXiv:2407.00459
  [hep-lat]}}.

\bibitem{Madore:1991bw}
J.~Madore, ``{The Fuzzy sphere},''
  \href{http://dx.doi.org/10.1088/0264-9381/9/1/008}{{\em Class. Quant. Grav.}
  {\bfseries 9} (1992) 69--88}.

\bibitem{Nekrasov:1998ss}
N.~Nekrasov and A.~S. Schwarz, ``{Instantons on noncommutative R**4 and (2,0)
  superconformal six-dimensional theory},''
  \href{http://dx.doi.org/10.1007/s002200050490}{{\em Commun. Math. Phys.}
  {\bfseries 198} (1998) 689--703},
  \href{http://arxiv.org/abs/hep-th/9802068}{{\ttfamily arXiv:hep-th/9802068}}.

\bibitem{Gubser:2000cd}
S.~S. Gubser and S.~L. Sondhi, ``{Phase structure of noncommutative scalar
  field theories},''
  \href{http://dx.doi.org/10.1016/S0550-3213(01)00108-0}{{\em Nucl. Phys. B}
  {\bfseries 605} (2001) 395--424},
  \href{http://arxiv.org/abs/hep-th/0006119}{{\ttfamily arXiv:hep-th/0006119}}.

\bibitem{Douglas:2001ba}
M.~R. Douglas and N.~A. Nekrasov, ``{Noncommutative field theory},''
  \href{http://dx.doi.org/10.1103/RevModPhys.73.977}{{\em Rev. Mod. Phys.}
  {\bfseries 73} (2001) 977--1029},
  \href{http://arxiv.org/abs/hep-th/0106048}{{\ttfamily arXiv:hep-th/0106048}}.

\bibitem{Szabo:2001kg}
R.~J. Szabo, ``{Quantum field theory on noncommutative spaces},''
  \href{http://dx.doi.org/10.1016/S0370-1573(03)00059-0}{{\em Phys. Rept.}
  {\bfseries 378} (2003) 207--299},
  \href{http://arxiv.org/abs/hep-th/0109162}{{\ttfamily arXiv:hep-th/0109162}}.

\bibitem{Seiberg:1999vs}
N.~Seiberg and E.~Witten, ``{String theory and noncommutative geometry},''
  \href{http://dx.doi.org/10.1088/1126-6708/1999/09/032}{{\em JHEP} {\bfseries
  09} (1999) 032}, \href{http://arxiv.org/abs/hep-th/9908142}{{\ttfamily
  arXiv:hep-th/9908142}}.

\bibitem{Minwalla:1999px}
S.~Minwalla, M.~Van~Raamsdonk, and N.~Seiberg, ``{Noncommutative perturbative
  dynamics},'' \href{http://dx.doi.org/10.1088/1126-6708/2000/02/020}{{\em
  JHEP} {\bfseries 02} (2000) 020},
  \href{http://arxiv.org/abs/hep-th/9912072}{{\ttfamily arXiv:hep-th/9912072}}.

\bibitem{Matusis:2000jf}
A.~Matusis, L.~Susskind, and N.~Toumbas, ``{The IR / UV connection in the
  noncommutative gauge theories},''
  \href{http://dx.doi.org/10.1088/1126-6708/2000/12/002}{{\em JHEP} {\bfseries
  12} (2000) 002}, \href{http://arxiv.org/abs/hep-th/0002075}{{\ttfamily
  arXiv:hep-th/0002075}}.

\bibitem{Chu:2001xi}
C.-S. Chu, J.~Madore, and H.~Steinacker, ``{Scaling limits of the fuzzy sphere
  at one loop},'' \href{http://dx.doi.org/10.1088/1126-6708/2001/08/038}{{\em
  JHEP} {\bfseries 08} (2001) 038},
  \href{http://arxiv.org/abs/hep-th/0106205}{{\ttfamily arXiv:hep-th/0106205}}.

\bibitem{Vaidya:2003ew}
S.~Vaidya and B.~Ydri, ``{On the origin of the UV-IR mixing in noncommutative
  matrix geometry},''
  \href{http://dx.doi.org/10.1016/j.nuclphysb.2003.08.023}{{\em Nucl. Phys. B}
  {\bfseries 671} (2003) 401--431},
  \href{http://arxiv.org/abs/hep-th/0305201}{{\ttfamily arXiv:hep-th/0305201}}.

\bibitem{Zhou:2023qfi}
Z.~Zhou, L.~Hu, W.~Zhu, and Y.-C. He, ``{SO(5) Deconfined Phase Transition
  under the Fuzzy-Sphere Microscope: Approximate Conformal Symmetry,
  Pseudo-Criticality, and Operator Spectrum},''
  \href{http://dx.doi.org/10.1103/PhysRevX.14.021044}{{\em Phys. Rev. X}
  {\bfseries 14} no.~2, (2024) 021044},
  \href{http://arxiv.org/abs/2306.16435}{{\ttfamily arXiv:2306.16435
  [cond-mat.str-el]}}.

\bibitem{Zou:2019dnc}
Y.~Zou, A.~Milsted, and G.~Vidal, ``{Conformal fields and operator product
  expansion in critical quantum spin chains},''
  \href{http://dx.doi.org/10.1103/PhysRevLett.124.040604}{{\em Phys. Rev.
  Lett.} {\bfseries 124} no.~4, (2020) 040604},
  \href{http://arxiv.org/abs/1901.06439}{{\ttfamily arXiv:1901.06439
  [cond-mat.str-el]}}.

\bibitem{Hu:2023ghk}
L.~Hu, Y.-C. He, and W.~Zhu, ``{Solving conformal defects in 3D conformal field
  theory using fuzzy sphere regularization},''
  \href{http://dx.doi.org/10.1038/s41467-024-47978-y}{{\em Nature Commun.}
  {\bfseries 15} no.~1, (2024) 3659},
  \href{http://arxiv.org/abs/2308.01903}{{\ttfamily arXiv:2308.01903
  [cond-mat.stat-mech]}}.

\bibitem{Zhou:2023fqu}
Z.~Zhou, D.~Gaiotto, Y.-C. He, and Y.~Zou, ``{The $g$-function and Defect
  Changing Operators from Wavefunction Overlap on a Fuzzy Sphere},''
  \href{http://arxiv.org/abs/2401.00039}{{\ttfamily arXiv:2401.00039
  [hep-th]}}.

\bibitem{Cuomo:2024psk}
G.~Cuomo, Y.-C. He, and Z.~Komargodski, ``{Impurities with a cusp: general
  theory and 3d Ising},'' \href{http://arxiv.org/abs/2406.10186}{{\ttfamily
  arXiv:2406.10186 [hep-th]}}.

\bibitem{Hu:2024pen}
L.~Hu, W.~Zhu, and Y.-C. He, ``{Entropic $F$-function of 3D Ising conformal
  field theory via the fuzzy sphere regularization},''
  \href{http://arxiv.org/abs/2401.17362}{{\ttfamily arXiv:2401.17362
  [hep-th]}}.

\bibitem{Chen:2024jxe}
B.-B. Chen, X.~Zhang, and Z.~Y. Meng, ``{Emergent Conformal Symmetry at the
  Multicritical Point of (2+1)D SO(5) Model with Wess-Zumino-Witten Term on
  Sphere},'' \href{http://arxiv.org/abs/2405.04470}{{\ttfamily arXiv:2405.04470
  [cond-mat.str-el]}}.

\bibitem{Ded:fuzzy}
M.~Dedushenko, ``{Notes on the fuzzy sphere and classical limit},'' {\em to
  appear} .

\bibitem{Cardy:1996xt}
J.~L. Cardy, {\em {Scaling and renormalization in statistical physics}}.
\newblock Cambridge University Press, 1996.

\bibitem{Diehl:1986}
H.~Diehl, {\em {Phase transitions and critical phenomena}}, vol.~10.
\newblock Academic Press, 1986.

\bibitem{Diehl:1998mh}
H.~W. Diehl and M.~Shpot, ``{Massive field theory approach to surface critical
  behavior in three-dimensional systems},''
  \href{http://dx.doi.org/10.1016/S0550-3213(98)00489-1}{{\em Nucl. Phys. B}
  {\bfseries 528} (1998) 595--647},
  \href{http://arxiv.org/abs/cond-mat/9804083}{{\ttfamily
  arXiv:cond-mat/9804083}}.

\bibitem{Liendo:2012hy}
P.~Liendo, L.~Rastelli, and B.~C. van Rees, ``{The Bootstrap Program for
  Boundary CFT$_d$},'' \href{http://dx.doi.org/10.1007/JHEP07(2013)113}{{\em
  JHEP} {\bfseries 07} (2013) 113},
  \href{http://arxiv.org/abs/1210.4258}{{\ttfamily arXiv:1210.4258 [hep-th]}}.

\bibitem{Gliozzi:2015qsa}
F.~Gliozzi, P.~Liendo, M.~Meineri, and A.~Rago, ``{Boundary and Interface CFTs
  from the Conformal Bootstrap},''
  \href{http://dx.doi.org/10.1007/JHEP05(2015)036}{{\em JHEP} {\bfseries 05}
  (2015) 036}, \href{http://arxiv.org/abs/1502.07217}{{\ttfamily
  arXiv:1502.07217 [hep-th]}}.

\bibitem{Metlitski:2020cqy}
M.~A. Metlitski, ``{Boundary criticality of the O(N) model in d = 3 critically
  revisited},'' \href{http://dx.doi.org/10.21468/SciPostPhys.12.4.131}{{\em
  SciPost Phys.} {\bfseries 12} no.~4, (2022) 131},
  \href{http://arxiv.org/abs/2009.05119}{{\ttfamily arXiv:2009.05119
  [cond-mat.str-el]}}.

\bibitem{Padayasi:2021sik}
J.~Padayasi, A.~Krishnan, M.~A. Metlitski, I.~A. Gruzberg, and M.~Meineri,
  ``{The extraordinary boundary transition in the 3d O(N) model via conformal
  bootstrap},'' \href{http://dx.doi.org/10.21468/SciPostPhys.12.6.190}{{\em
  SciPost Phys.} {\bfseries 12} no.~6, (2022) 190},
  \href{http://arxiv.org/abs/2111.03071}{{\ttfamily arXiv:2111.03071
  [cond-mat.stat-mech]}}.

\bibitem{Toldin:2021kun}
F.~P. Toldin and M.~A. Metlitski, ``{Boundary Criticality of the 3D O(N) Model:
  From Normal to Extraordinary},''
  \href{http://dx.doi.org/10.1103/PhysRevLett.128.215701}{{\em Phys. Rev.
  Lett.} {\bfseries 128} no.~21, (2022) 215701},
  \href{http://arxiv.org/abs/2111.03613}{{\ttfamily arXiv:2111.03613
  [cond-mat.stat-mech]}}.

\bibitem{Trepanier:2023tvb}
M.~Tr\'epanier, ``{Surface defects in the O(N) model},''
  \href{http://dx.doi.org/10.1007/JHEP09(2023)074}{{\em JHEP} {\bfseries 09}
  (2023) 074}, \href{http://arxiv.org/abs/2305.10486}{{\ttfamily
  arXiv:2305.10486 [hep-th]}}.

\bibitem{Assel:2015nca}
B.~Assel, D.~Cassani, L.~Di~Pietro, Z.~Komargodski, J.~Lorenzen, and
  D.~Martelli, ``{The Casimir Energy in Curved Space and its Supersymmetric
  Counterpart},'' \href{http://dx.doi.org/10.1007/JHEP07(2015)043}{{\em JHEP}
  {\bfseries 07} (2015) 043}, \href{http://arxiv.org/abs/1503.05537}{{\ttfamily
  arXiv:1503.05537 [hep-th]}}.

\bibitem{Luo:2022tqy}
C.~Luo and Y.~Wang, ``{Casimir energy and modularity in higher-dimensional
  conformal field theories},''
  \href{http://dx.doi.org/10.1007/JHEP07(2023)028}{{\em JHEP} {\bfseries 07}
  (2023) 028}, \href{http://arxiv.org/abs/2212.14866}{{\ttfamily
  arXiv:2212.14866 [hep-th]}}.

\bibitem{Polyakov:1970xd}
A.~M. Polyakov, ``{Conformal symmetry of critical fluctuations},'' {\em JETP
  Lett.} {\bfseries 12} (1970) 381--383.

\bibitem{Wu:1976ge}
T.~T. Wu and C.~N. Yang, ``{Dirac Monopole Without Strings: Monopole
  Harmonics},'' \href{http://dx.doi.org/10.1016/0550-3213(76)90143-7}{{\em
  Nucl. Phys. B} {\bfseries 107} (1976) 365}.

\bibitem{MaMarinescu-book}
X.~Ma and G.~Marinescu, {\em {Holomorphic Morse Inequalities and Bergman
  Kernels}}.
\newblock Birkh\, 2007.

\bibitem{MaMarinescu+2012+1+56}
X.~Ma and G.~Marinescu, ``{Berezin-Toeplitz quantization on Khler manifolds},''
  \href{http://dx.doi.org/doi:10.1515/CRELLE.2011.133}{{\em Journal f\"{u}r die
  reine und angewandte Mathematik} {\bfseries 2012} no.~662, (2012) 1--56},
  \href{http://arxiv.org/abs/1009.4405}{{\ttfamily arXiv:1009.4405 [math.DG]}}.

\bibitem{schlichenmaier2010berezin}
M.~Schlichenmaier, ``{Berezin-Toeplitz Quantization for Compact K{\"a}hler
  Manifolds. A Review of Results},''
  \href{http://dx.doi.org/10.1155/2010/927280}{{\em Advances in Mathematical
  Physics} {\bfseries 2010} no.~1, (2010) 927280}.

\bibitem{greiter2011landau}
M.~Greiter, ``{Landau level quantization on the sphere},'' {\em Physical Review
  B} {\bfseries 83} no.~11, (2011) 115--129,
  \href{http://arxiv.org/abs/1101.3943}{{\ttfamily arXiv:1101.3943
  [cond-mat.str-el]}}.

\bibitem{Ippoliti:2018ojo}
M.~Ippoliti, R.~S.~K. Mong, F.~F. Assaad, and M.~P. Zaletel, ``{Half-filled
  Landau levels: A continuum and sign-free regularization for three-dimensional
  quantum critical points},''
  \href{http://dx.doi.org/10.1103/PhysRevB.98.235108}{{\em Phys. Rev. B}
  {\bfseries 98} no.~23, (2018) 235108},
  \href{http://arxiv.org/abs/1810.00009}{{\ttfamily arXiv:1810.00009
  [cond-mat.str-el]}}.

\bibitem{Haldane:1983xm}
F.~D.~M. Haldane, ``{Fractional quantization of the Hall effect: A Hierarchy of
  incompressible quantum fluid states},''
  \href{http://dx.doi.org/10.1103/PhysRevLett.51.605}{{\em Phys. Rev. Lett.}
  {\bfseries 51} (1983) 605--608}.

\bibitem{hasenbusch2011thermodynamic}
M.~Hasenbusch, ``{Thermodynamic Casimir force: A Monte Carlo study of the
  crossover between the ordinary and the normal surface universality class},''
  {\em Physical Review B -- Condensed Matter and Materials Physics} {\bfseries
  83} no.~13, (2011) 134425, \href{http://arxiv.org/abs/1012.4986}{{\ttfamily
  arXiv:1012.4986 [cond-mat.stat-mech]}}.

\end{thebibliography}\endgroup
\end{document}